\documentclass[aip,jmp,amsmath, amssymb, 
preprint, 
]{revtex4-1}
\usepackage{graphicx}
\usepackage{xcolor}
\usepackage[hypertexnames=false]{hyperref}
\hypersetup{
    colorlinks,
    linkcolor={black!50!black},
    citecolor={black!50!black},
    urlcolor={blue!80!black}
}

\begin{document} %%%%%%%%%%%%%%%%%%%%%%%%%%%%%%%%%%%%%%%%%%%%%%%%%%%%%%
\raggedbottom % do not equalize page contents

% Title
\title{Continuum models of the electrochemical diffuse layer in electronic-structure calculations}

% Authors 
\author{Francesco Nattino}
\affiliation{Theory and Simulations of Materials (THEOS) and National Centre for Computational Design 
and Discovery of Novel Materials (MARVEL), \'{E}cole Polytechnique F\'{e}d\'{e}rale de Lausanne, 
CH-1015 Lausanne, Switzerland.}

\author{Matthew Truscott}
\affiliation{Department of Physics, University of North Texas, Denton, TX 76207, USA.}

\author{Nicola Marzari}
\affiliation{Theory and Simulations of Materials (THEOS) and National Centre for Computational Design 
and Discovery of Novel Materials (MARVEL), \'{E}cole Polytechnique F\'{e}d\'{e}rale de Lausanne, 
CH-1015 Lausanne, Switzerland.}

\author{Oliviero Andreussi}
\affiliation{Department of Physics, University of North Texas, Denton, TX 76207, USA.}

% Date 
\date{\today} 

%%%%%%%%%%%%%%%%%%%%%%%%%%%%%%%%%%%%%%%%%%%%%%%%%%%%%%%%%%%%%%%%%%%%%%
\begin{abstract}
Continuum electrolyte models represent a practical tool to account for the presence of 
the diffuse layer at electrochemical interfaces. However, despite the increasing popularity of these 
in the field of materials science it remains unclear which features 
are necessary in order to accurately describe interface-related observables such as the differential
capacitance (DC) of metal electrode surfaces. 
% elements of the available diffuse-layer models are 
%necessary in order to reproduce experimentally-measured observables such as the differential
%capacitance (DC) of single-crystal metal electrode surfaces. 
We present here a critical comparison of continuum diffuse-layer models that can be coupled to an
atomistic first-principles description of the charged metal surface in order to account for the electrolyte screening at electrified interfaces.
By comparing computed DC values for the 
prototypical Ag(100) surface in an aqueous solution to experimental data we validate the accuracy of the models considered.
Results suggest that a size-modified Poisson-Boltzmann description of the electrolyte solution is sufficient to 
qualitatively reproduce the main experimental trends. Our findings also highlight the large effect that the dielectric cavity parameterization has 
on the computed DC values. 
\end{abstract}

%%%%%%%%%%%%%%%%%%%%%%%%%%%%%%%%%%%%%%%%%%%%%%%%%%%%%%%%%%%%%%%%%%%%%%
\maketitle

%%%%%%%%%%%%%%%%%%%%%%%%%%%%%%%%%%%%%%%%%%%%%%%%%%%%%%%%%%%%%%%%%%%%%%
\section{INTRODUCTION\label{sec:Introduction}}

The electrical double layer (DL) is of primary importance in the field of energy conversion,
as it plays a crucial role in devices such as supercapacitors and fuel cells\cite{Simon2008, Biesheuvel2009}.
The DL structure is essentially characterized
by two layers of opposite charge that appear at the interface between an
electrified surface and an electrolyte solution. This structure arises from the charge accumulation 
at the boundary of the solvated surface, which attracts counterions from the bulk solution.
The balance between the electrostatic attraction towards the charged surface,
the entropic electrolyte contributions, and the steric repulsion between the ions 
gives rise to an equilibrium charge  
distribution in the solution that is generally known as the diffuse layer. 

Unfortunately, 
various limitations hamper
atomistic simulations of the diffuse layer\cite{Yeh2013}. 
First, long simulation times are required in order to achieve statistically significant samplings of the solvent 
and electrolyte configurations, with the corresponding time-scales being often beyond the reach of standard first-principles
molecular dynamics techniques. In addition, large simulation cells are necessary in order
 to capture the long-range screening of typical values of the surface charge densities. 

\begin{figure}%++++++++++++++++++++++++++++++++++++++++++++++++++++++
\begin{centering}
	\includegraphics[width=0.8\columnwidth]{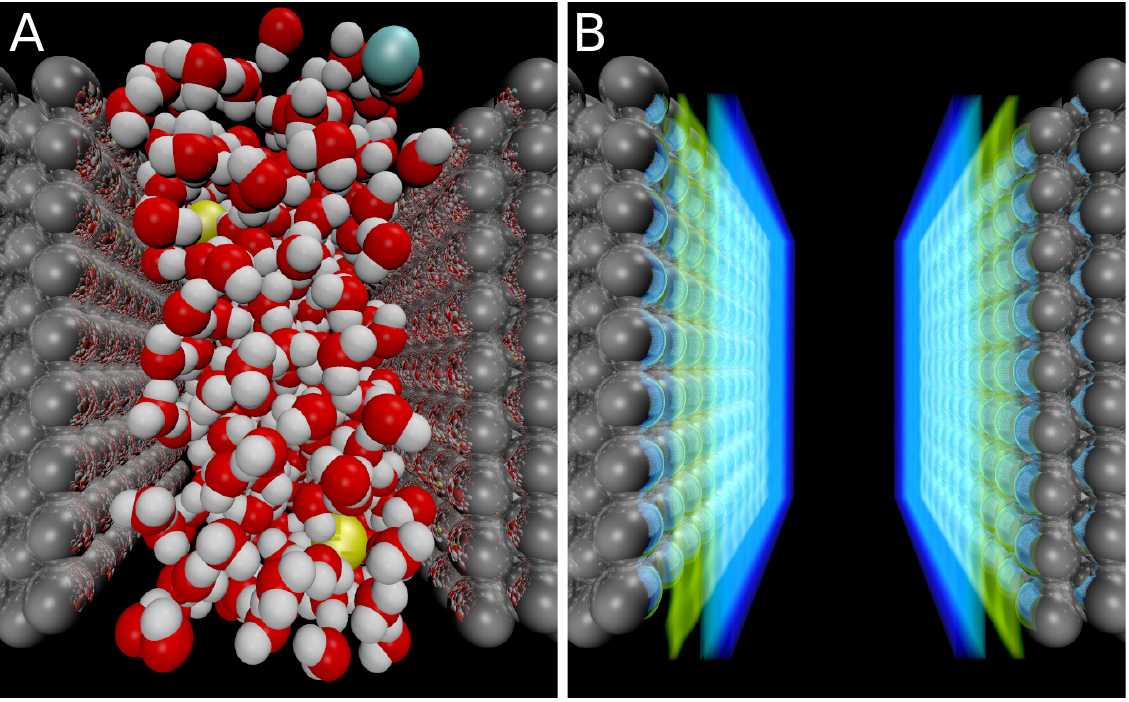}\\
\end{centering}
\caption{
Visualization of the atomistic details of a typical electrochemical setup (A): The metal slab (silver spheres) is in contact with an neutral aqueous solution (oxygen in red, hydrogen in white) containing electrolyte species (cyan and yellow spheres). Continuum models are obtained by integrating out the atomistic degrees of freedom of the mobile species (e.g. water and electrolytes) and replacing them with homogeneous continuum bodies, whose boundaries reflect the physical separation between the QM system and the environment. In B the solvent  boundary (green transparent surface) is reported together with the continuous charge density of the electrolyte (blue transparent field), as computed for a charged substrate. Different onset for the solvent and electrolyte continuum models can be imposed in the definition of the model or can be obtained by including additional repulsive interactions between the continuum electrolyte and the QM substrate.}
\label{fig:continuum}
\end{figure}

Continuum models represent an attractive alternative to fully-atomistic models of electrolyte solutions. A 
continuum description of the solvent and of the ions allows, in fact, to bypass the computationally-intensive 
configurational sampling of the solution's degrees of freedom. 
In particular, our focus here is on hybrid methods, where a first-principles modeling of an electrified surface 
is coupled to a continuum description of the solution (Figure \ref{fig:continuum}). These models are
particularly appealing for the accuracy and predictive power that they 
can potentially have, as the processes occurring at or within the metal 
surface are described at a quantum-mechanical level, while the electrostatic screening of the
diffuse layer is accounted for at a  mean-field level. 

%+++++++++++++++++++++++++++++++++++++++++++++++++++++++

Starting from highly simplified models of the double layer,
which consist of a countercharge plane at a fixed distance from a charged metal surface\cite{Fu1989, Lozovoi2003, Bonnet2013First-principlesConditions},
more complex diffuse layer models have been subsequently proposed and integrated into
periodic density-functional theory (DFT) codes\cite{otani_first-principles_2006, Jinnouchi-PRB-2008,
DaboThesis, Dabo-arXiv-2008, Dabo2010, LetchworthWeaver-PRB-2012, Gunceler2013TheSystems, Mathew2016ImplicitTheory,
Sundararaman2017EvaluatingImprovement, Sundararaman2018ImprovingCalculations, Melander2018}. 
However, despite the large variety of electrolyte models proposed, there is no 
consensus on the model features required to achieve a physically-sound
description of the diffuse layer. On one hand, full-continuum models 
that are based on the solution of some form of the size-modified Poisson-Boltzmann (PB)
equation have been shown to qualitatively or semi-quantitatively describe experimental data\cite{Bazant2009, Nakayama2015, Baskin-JElectrochemSoc-2017}. 
On the other hand, recent work from Sundararaman et al.\cite{Sundararaman2018ImprovingCalculations} has suggested that non-linear 
effects in the dielectric continuum also play an important role, and should thus be 
accounted for in order to reproduce measured trends.
%Specifically, most of the available 
%electrolyte models are now based on the solution of some form of the Poisson-Boltzmann (PB)
%equation, which accounts for the equilibrium between electrostatic forces and thermal motion
%of the ions. %However, different flavors of the PB model have been proposed so far. In addition, 
%it is often difficult to disentangle the role of the electrolyte from the one of the dielectric continuum  
%that is generally employed to mimic the solvent screening. 

In this work, we tackle these issues by systematically analyzing the 
performance of a hierarchy of continuum diffuse layer models
of increasing complexity. 
We show how the various models can be derived from similar
expressions of a free-energy functional and how they can be implemented 
in the framework of DFT. We choose the differential capacitance (DC) of a model 
metal surface as a prototypical observable to compare and contrast the 
various electrolyte models. In particular, we consider a Ag(100) surface 
in an aqueous solution as study system, motivated by the availability of accurate
experimental data \cite{Valette-JElectroanalChem-1982} 
that have been widely used in the literature to validate 
diffuse layer models \cite{Bazant2009, Sundararaman2017EvaluatingImprovement, Sundararaman2018ImprovingCalculations}.

Results show that a size-modified Poisson-Boltzmann model is 
able to qualitatively capture the main features of experimental DC curves, including
the minimum capacitance value at the potential of zero charge, and the two 
local maxima at higher and lower potentials. The choice of solvation cavity employed
to separate the 
quantum-mechanical region from the continuum solvent region is also found to play an 
important role on the absolute value of the computed DC. 

The article is structured as follows. Section \ref{sec:Methods} reviews the theoretical
background of the diffuse layer models considered and it presents the details of their computational 
implementations. Results on the computed DC values for the Ag(100) surface are then presented 
in Section \ref{sec:Results-and-Discussion}. In particular, the various electrolyte models 
are illustratively compared under vacuum conditions in Section \ref{sub:results-vacuum}, while their 
performance is better validated in Section \ref{sub:results-solvent} by comparing results to experimental data.
Finally, the conclusions are presented in Section \ref{sec:Summary-and-Conclusions}.

%%%%%%%%%%%%%%%%%%%%%%%%%%%%%%%%%%%%%%%%%%%%%%%%%%%%%%%%%%%%%%%%%%%%%%
\section{METHODS\label{sec:Methods}}

\subsection{The Electrolyte Cavity\label{sub:cavity}}

In the framework of continuum solvation models, the 
solvent's degrees of freedoms are smeared out in a 
continuum description and accounted for 
by means of a dielectric continuum. 
An important element in this class of models is represented 
by the so-called solvation cavity, which defines the 
boundary between the explicitly described solute region and the 
solvent region, where the dielectric
continuum is located. 
This partitioning of the simulation cell can be formally defined through an interface function, 
$s(\textbf{r})$. We define here $s(\textbf{r}) \equiv 1$ inside the solute region, 
and $s(\textbf{r}) \equiv 0$ in the region of space characterized by the solvent dielectric constant $\varepsilon_0$. 
In the field of material science and condensed matter physics, continuum models
 typically involve interface functions that smoothly switch between the two regions,
 as they turn out to provide a considerable improvement to numerical stability\cite{Fattebert2003First-principlesSolvent, scherlis_jcp_2006, Andreussi-JCP-2012}. 
Closely related to $s(\textbf{r})$, the dielectric function $\varepsilon(\textbf{r})$
sets the local value of the dielectric constant:
\begin{equation}
\varepsilon\left(\mathbf{r}\right) = 1 + \left(\varepsilon_0 - 1\right)\left(1 - s(\textbf{r})\right). \label{eq:dielectric}
\end{equation}

In a similar fashion, the interface function $s(\textbf{r})$ can be exploited to define
the region of space that is accessible to the ionic species in the solution. In particular,
the complementary interface function $\gamma\left(\mathbf{r}\right)$ defines 
the portion of the cell where the electrolyte solution is located:
\begin{equation}
\gamma\left(\mathbf{r}\right) = 1 -  s\left(\textbf{r}\right). \label{eq:gamma}
\end{equation}

It is important to stress here that in the above equations we have expressed both the solvent and the electrolyte domains in terms of the same interface function. In principle, since the two domains are associated with the regions of space that are accessible to solvent and electrolyte particles, respectively, different interface functions should be required. For example, electrolyte ions that have strong solvation shells may be hindered direct access to the electrochemical interface, their closest distance from the substrate being increased by the thickness of the coordinating solvent molecules in what is known as the Stern layer\cite{Stern1924}. In order to simplify the parameterization and tuning of the different interfaces, the electrolyte boundary is often expressed as a scaled version of the solvent one\cite{Dabo-arXiv-2008, Harris2014, Ringe2016Function-Space-BasedDFT, Sundararaman2018ImprovingCalculations}. Alternatively, a single interface is used and additional repulsive potentials are added to the free energy functional of the electrolyte system to stabilize solutions which are displaced from the solvent interface\cite{Jinnouchi-PRB-2008}. The latter approach has the additional flexibility of allowing both repulsive and attractive interactions between the components of the diffuse layer and the substrate. For this reason, we decided to focus the following discussion on models with a single common interface function. 

The interface function $s(\textbf{r})$ is typically constructed as a function of the 
solute's degrees of freedom.  
In order to explore how the choice of the solvation cavity affects the 
resulting diffuse layer model, we have considered the following 
three interface functions.
 
A first interface function is based on the local value of 
the solute's electron density. An optimally-smooth switching function
has been proposed by Andreussi et al.\cite{Andreussi-JCP-2012}, 
who have revised in the so-called self-consistent continuum solvation (SCCS) 
model the one originally proposed by Fattebert and Gigy\cite{Fattebert2003First-principlesSolvent}
and Scherlis et al. \cite{scherlis_jcp_2006}. 

As a second interface function, we consider a rigid function 
of the solute's atomic positions, as defined through 
the product of atom-centered interlocking spheres with a smooth
erf-like profile. This cavity function, as proposed by Fisicaro et al.
in the recent soft-sphere continuum solvation (SSCS) model\cite{Fisicaro2017},  
accounts for the diversity of the chemical species involved through
tabulated van-der-Waals radii.

The last interface function reflects the two-dimensional character of 
the slab system considered. In particular, a planar boundary between solvent and solute 
is defined as a mere function of the vertical distance from the slab center, $d$. 
Two parameters are employed to define the cavity.
The first one, $d$, defines the absolute position of the interface. The second one, $\Delta$, regulates 
the smoothness of the boundary by defining the spread
of the error-function profile which we have included along the surface normal. This simplified form of interface has not been explicitly addressed before in the literature. It has been introduced here to facilitate the comparison with analytical one-dimensional models. For particularly regular substrates it may provide an easier and more robust approach with respect to atomic-centered or electron-based interfaces.
 
%%%%
%While in principle different interfaces 
%could be set for solvent and electrolyte in order to introduce solvent-accessible but 
%ion-excluded regions (i.e. the Stern layer), the role 
%of the interface function in the framework of implicit solvation is to 
%set the boundary between quantum-mechanical and continuum regions. 
%We thus find it more appropriate to use the same interface function
% $s(\textbf{r})$ for both electrolyte and solvent. One can still account for 
% an ion-free region in correspondence of the solute first solvation shell(s)
% by introducing, for instance, an \emph{ad-hoc} repulsive interaction 
% between the solute and the electrolyte species, as described below. 
%%%% 

\subsection{The Electrolyte Models\label{sub:electrolyte-models}}

In this section we describe the continuum electrolyte 
models considered and illustrate how they can be derived from specific free-energy functionals
where we include all electrostatic and mean-field contributions (the 
usual non-interacting electron kinetic energy and the exchange-correlation 
energy terms are also included, but left out in the expressions for improved readability). 
In order to perform self-consistent DFT calculations for a system embedded in 
an electrolyte solution, energy contributions that explicitly depend on the 
solute electron density require the inclusion of corresponding 
terms in the Kohn-Sham (KS) potential. Furthermore, terms that explicitly depend
on the solute's atomic positions give rise to analogous contributions to the 
atomic forces. All these contributions are reported in the 
Supplemental Material. 

In this work, we have neglected the solvent-related 
non-electrostatic contributions to the free-energy\cite{scherlis_jcp_2006, Andreussi-JCP-2012}
based on the quantum volume and quantum surface\cite{cococcioni_prl_2005}. 
Such contributions, however, can be straightforwardly included by adding corresponding terms to
 the free-energy, to the KS potential and to the forces\cite{scherlis_jcp_2006, Andreussi-JCP-2012}.

\subsubsection{Planar Countercharge Model\label{subsub:helmholtz}}

To first approximation, the electrolyte screening of the surface charge
can be accounted for by introducing a countercharge plane 
at a given distance from the surface\cite{Fu1989}. 
The presence of this external charge modifies the electrostatic energy 
of the system. When accounting for this electrostatic term, 
the free energy of the system embedded
in the electrolyte solution can be computed as:
%\begin{multline}
%F^{PC}\left[\rho\left(\mathbf{r}\right),\phi\left(\mathbf{r}\right) \right] = \int\left[-\frac{\varepsilon\left(\mathbf{r}\right)}{8\pi}\left|\nabla\phi\left(\mathbf{r}\right)\right|^{2}+\rho\left(\mathbf{r}\right)\phi\left(\mathbf{r}\right)+\right.\\
%\left.\rho^{ions}\left(\mathbf{r}\right)\phi\left(\mathbf{r}\right)\right]d\mathbf{r}\label{eq:free-energy-helmholtz}
%\end{multline}
\begin{equation}
F^{PC}\left[\rho\left(\mathbf{r}\right),\phi\left(\mathbf{r}\right) \right] = \int\left[-\frac{\varepsilon\left(\mathbf{r}\right)}{8\pi}\left|\nabla\phi\left(\mathbf{r}\right)\right|^{2}+\rho\left(\mathbf{r}\right)\phi\left(\mathbf{r}\right)+\rho^{ions}\left(\mathbf{r}\right)\phi\left(\mathbf{r}\right)\right]d\mathbf{r}\label{eq:free-energy-helmholtz}
\end{equation}
Here $\rho\left(\mathbf{r}\right)$ is the total (i.e. electronic + nuclear) 
charge density of the solute,  $\phi\left(\mathbf{r}\right)$ is the
electrostatic potential and  
$\rho^{ions}\left(\mathbf{r}\right)$ is the external charge density
that mimics the counterion accumulation.
 
This model can be seen as a computational 
implementation of the Helmholtz model for the double layer\cite{Helmholtz1853}: 
the countercharge plane completely screens the 
surface charge in a
region of space that can be chosen to be infinitely narrow.
Note that the Helmholtz screening does not depend on the ionic
strength of the solution; it is thus not surprising that the 
bulk electrolyte concentration does not appear in Eq. \ref{eq:free-energy-helmholtz}.

\subsubsection{Poisson-Boltzmann Model\label{subsub:PB}}

A more physical description of the diffuse layer 
 can be derived from a free-energy
expression that accounts for the chemical 
potential and the entropy of the ions in the solution\cite{Borukhov-PRL-1997, Dabo-arXiv-2008}. 
These terms allow one to introduce an explicit dependence on 
the local concentrations of the electrolyte species
($\{c_{i}\left(\mathbf{r}\right)\}$).
For an electrolyte
solution with $p$ ionic species with charges $\{z_i\}$ and bulk
concentrations $\{c^0_i\}$, such that the solution is overall neutral 
($\sum_{i=1}^p z_i c^0_i = 0$), 
the free energy functional takes the following form\cite{Borukhov-PRL-1997, Dabo-arXiv-2008}:
%\begin{multline}
%F\left[\rho\left(\mathbf{r}\right),\phi\left(\mathbf{r}\right),\left\{ c_{i}\left(\mathbf{r}\right)\right\} \right] = \int\left[-\frac{\varepsilon\left(\mathbf{r}\right)}{8\pi}\left|\nabla\phi\left(\mathbf{r}\right)\right|^{2}+\right.\\
%+\rho\left(\mathbf{r}\right)\phi\left(\mathbf{r}\right)+\rho^{ions}\left(\mathbf{r}\right)\phi\left(\mathbf{r}\right) -\sum_{i=1}^p\mu_{i}\left(c_{i}\left(\mathbf{r}\right)-c_{i}^{0}\right)+
% \\\left.-T\left(s\left[\left\{ c_{i}\left(\mathbf{r}\right)\right\} \right]-s\left[\left\{ c_{i}^{0}\right\} \right]\right)\right]\mathrm{d}\mathbf{r}.\label{eq:free-energy-PB}
%\end{multline}
\begin{multline}
F\left[\rho\left(\mathbf{r}\right),\phi\left(\mathbf{r}\right),\left\{ c_{i}\left(\mathbf{r}\right)\right\} \right] = \int\left[-\frac{\varepsilon\left(\mathbf{r}\right)}{8\pi}\left|\nabla\phi\left(\mathbf{r}\right)\right|^{2}+\rho\left(\mathbf{r}\right)\phi\left(\mathbf{r}\right)+\rho^{ions}\left(\mathbf{r}\right)\phi\left(\mathbf{r}\right) +\right.
 \\\left.-\sum_{i=1}^p\mu_{i}\left(c_{i}\left(\mathbf{r}\right)-c_{i}^{0}\right)-T\left(s\left[\left\{ c_{i}\left(\mathbf{r}\right)\right\} \right]-s\left[\left\{ c_{i}^{0}\right\} \right]\right)\right]\mathrm{d}\mathbf{r}.\label{eq:free-energy-PB}
\end{multline}
Here $\mu_i$ is the chemical potential of the $i$-th electrolyte species, 
$T$ is the temperature, and $s[\{c_i\}]$ is the electrolyte entropy 
density per unit volume.
The electrolyte charge density can be expressed in terms of the local electrolyte 
concentrations as
$\rho^{ions}\left(\mathbf{r}\right) = \sum_{i=1}^{p}c_{i}\left(\mathbf{r}\right)z_{i}$.

Under the assumptions of a point-charge electrolyte and ideal mixing, the entropy density of the solution is:
\begin{align}
s\left[\left\{ c_{i}\left(\mathbf{r}\right)\right\} \right]=-k_{B}\sum_{i=1}^{p}c_{i}\left(\mathbf{r}\right)\ln\frac{c_{i}\left(\mathbf{r}\right)}{\gamma\left(\mathbf{r}\right)}, \label{eq:entropy-PB}
\end{align}
where $k_{B}$ is the Boltzmann constant.
Note that the exclusion function $\gamma\left(\mathbf{r}\right)$, which sets the boundary between the electrolyte solution and the solute region,
enforces a zero entropic contribution from the volume assigned to the quantum-mechanical region. 
%As also noted by Ringe et al.\ref{Ringe2016Function-Space-BasedDFT}, the same free-energy expression is recovered if one
%assumes the entire volume of the cell (including the solute region) as part of the
%electrolyte solution and accounts for a repulsive potential $\varphi\left(\mathbf{r}\right)$ between the solute
%and the ionic species, such that $\gamma\left(\mathbf{r}\right)\equiv e^{-\frac{\varphi\left(\mathbf{r}\right)}{k_BT}}$. 

In order to find an expression for the equilibrium electrolyte 
concentrations, the free-energy functional in equation 
\ref{eq:free-energy-PB} is minimized with respect to
$c_{i}\left(\mathbf{r}\right)$. This procedure first leads to the condition:
\begin{align}
z_{i}\phi\left(\mathbf{r}\right)-\mu_{i}+k_{B}T\left(\ln\frac{c_{i}\left(\mathbf{r}\right)}{\gamma\left(\mathbf{r}\right)} + 1 \right)=0, \label{eq:chemical-potential-condition}
\end{align}
which allows one to obtain an expression for the chemical potential from the 
 the bulk electrolyte region, where $\phi\left(\mathbf{r}\right) = 0$ and $\gamma\left(\mathbf{r}\right) = 1$, obtaining:
\begin{align}
\mu_{i}=k_{B}T\left(\ln c_{i}^{0} + 1\right).\label{eq:chemical-potential-PB}
\end{align}
By substituting equation \ref{eq:chemical-potential-PB} back into equation \ref{eq:chemical-potential-condition}
one then obtains the following expression for the equilibrium electrolyte concentration:
\begin{align}
c_{i}\left(\mathbf{r}\right)=\gamma\left(\mathbf{r}\right)c_{i}^{0}e^{-\frac{z_{i}\phi\left(\mathbf{r}\right)}{k_{B}T}} \equiv c^{PB}_{i}\left(\phi\left(\mathbf{r}\right)\right) \label{eq:concentration-PB}
\end{align}
By using this equilibrium electrolyte concentration, the free-energy 
functional expression significantly simplifies to: 
%\begin{multline}
%F^{PB}\left[\rho\left(\mathbf{r}\right),\phi\left(\mathbf{r}\right)\right] = \int\left[-\frac{\varepsilon\left(\mathbf{r}\right)}{8\pi}\left|\nabla\phi\left(\mathbf{r}\right)\right|^{2}+\rho\left(\mathbf{r}\right)\phi\left(\mathbf{r}\right) +\right.\\
%+\left. k_BT\sum_{i=1}^{p}c_{i}^{0} \left(1 - \gamma\left(\mathbf{r}\right)e^{-\frac{z_i\phi\left(\mathbf{r}\right)}{k_BT}}\right) \right]\mathrm{d}\mathbf{r}. \label{eq:free-energy-PB-final}
%\end{multline}
\begin{equation}
F^{PB}\left[\rho\left(\mathbf{r}\right),\phi\left(\mathbf{r}\right)\right] = \int\left[-\frac{\varepsilon\left(\mathbf{r}\right)}{8\pi}\left|\nabla\phi\left(\mathbf{r}\right)\right|^{2}+\rho\left(\mathbf{r}\right)\phi\left(\mathbf{r}\right) + k_BT\sum_{i=1}^{p}c_{i}^{0} \left(1 - \gamma\left(\mathbf{r}\right)e^{-\frac{z_i\phi\left(\mathbf{r}\right)}{k_BT}}\right) \right]\mathrm{d}\mathbf{r}. \label{eq:free-energy-PB-final}
\end{equation}
Minimization of the free-energy functional with respect to $\phi\left(\mathbf{r}\right)$ now
leads to the well-known Poisson-Boltzmann equation (PBE), which allows one to relate the 
equilibrium charge densities in the system to the the electrostatic potential of the system: 
\begin{align}
\nabla\cdot\varepsilon\left(\mathbf{r}\right)\nabla\phi\left(\mathbf{r}\right)+4\pi\sum_{i=1}^{p}z_{i}c^{PB}_{i}\left(\phi\left(\mathbf{r}\right)\right)=-4\pi\rho\left(\mathbf{r}\right). \label{eq:PBE}
\end{align}

%Unfortunately, solving the PBE is not a trivial task, as the local concentrations of the
%electrolyte species depend exponentially on $\phi\left(\mathbf{r}\right)$. 
For low electrostatic potentials, i.e. whenever $z_i\phi\left(\mathbf{r}\right) \ll k_B T $, 
one can approximate the exponential dependence on $\phi\left(\mathbf{r}\right)$ with a linear function:
 \begin{equation}
e^{-\frac{z_{i}\phi\left(\mathbf{r}\right)}{k_{B}T}} \approx 1-\frac{z_{i}\phi\left(\mathbf{r}\right)}{k_{B}T}. 
\end{equation}
The expression of the electrolyte concentrations thus reduces to:
\begin{equation}
c_{i}\left(\mathbf{r}\right)\approx\gamma\left(\mathbf{r}\right)c_{i}^{0}\left( 1 -\frac{z_{i}\phi\left(\mathbf{r}\right)}{k_{B}T} \right)\equiv c^{LPB}\left(\phi\left(\mathbf{r}\right)\right), \label{eq:concentration-LPB}
\end{equation}
which leads to the following linearized-version of the PBE (LPBE): 
\begin{equation}
\nabla\cdot\varepsilon\left(\mathbf{r}\right)\nabla\phi\left(\mathbf{r}\right)-k^2\gamma\left(\mathbf{r}\right)\phi\left(\mathbf{r}\right)=-4\pi\rho\left(\mathbf{r}\right). \label{eq:LPBE}
\end{equation}
The constant operator $k^2 = 4\pi\frac{\sum_{i=1}^{p}z_{i}^{2}c_{i}^{0}}{k_{B}T}$ 
is related to the Debye length $\lambda_D$ of the electrolyte solution: 
\begin{equation}
k^2 = \frac{\varepsilon_0}{\lambda_D^2}.
\end{equation}
The LPBE can be equivalently derived\cite{Ringe2016Function-Space-BasedDFT} by Taylor-expanding the exponential term
 in Eq. \ref{eq:free-energy-PB-final} up to second order in $\phi\left(\mathbf{r}\right)$, and by subsequently
minimizing the resulting energy functional, 
\begin{multline}
F^{LPB}\left[\rho\left(\mathbf{r}\right),\phi\left(\mathbf{r}\right) \right] = \int\left[-\frac{\varepsilon\left(\mathbf{r}\right)}{8\pi}\left|\nabla\phi\left(\mathbf{r}\right)\right|^{2}+\rho\left(\mathbf{r}\right)\phi\left(\mathbf{r}\right) +\right.\\
\left.-\frac{\sum_{i=1}^{p}z_{i}^{2}c_{i}^{0}}{2k_{B}T}\gamma\left(\mathbf{r}\right)\phi^2\left(\mathbf{r}\right)+ k_{B}T\sum_{i=1}^{p}c_{i}^{0}\left(1 - \gamma\left(\mathbf{r}\right)\right) \right]\mathrm{d}\mathbf{r}, \label{eq:free-energy-LPB}
\end{multline}
with respect to the electrostatic potential.

The linear-regime of the PBE is expected to hold
for a narrow potential range around the potential of zero charge (PZC). However, typical applications 
easily require the modeling of potential windows that extend for hundreds of mV, making it desirable 
to have efficient strategies to solve the full PB problem instead.
If the interface between the solute and the electrolyte is suitable to a two-dimensional approximation, 
one can tackle the full PB problem by taking advantage of the reduced dimensionality of the interface. 
In particular, one can integrate out the dimensions in the surface plane and exploit the 
analytical solution of the PBE in one dimension\cite{Dabo-arXiv-2008, DaboThesis, Dabo2010}.
In the following, we assume for convenience that the system is oriented with the $x$ axis perpendicular 
to the slab plane and that the diffuse layer starts at a distance $x_{Stern}$ from the center of the slab. 
Taking the planar average of the physical quantities involved in Eq. \ref{eq:PBE} and assuming that the 
system charge density and the dielectric interfaces are fully contained within the $x_{Stern}$ distance, 
the resulting one-dimensional differential equation
\begin{equation}
\frac{d^2\phi\left(x\right)}{dx^2}=-\frac{4\pi} {\varepsilon_0}\left(\rho\left(x\right)+\sum_{i=1}^{p}z_ic_i\left(\phi\left(x\right)\right)\right)
\end{equation}
can be integrated analytically for $\left|x\right|\ge x_{Stern}$. For the most common case of a 
diffuse layer composed by ions of equal concentrations $c^0$ and opposite signs, the electrostatic potential 
in the electrolyte region can be expressed as (see Supplemental material for the derivation):
\begin{equation}
\phi\left(x\right) =\frac{4k_BT}{|z|} {\coth}^{-1}\left(c_1\exp\left(c_2\left|x\right|\right)\right),
\label{eq:GCS_potential}\end{equation}
where $c_2 = 32\pi k_BT c^0/\varepsilon_0$ and $c_1$ is obtained by imposing continuity of the normal 
component of the electric field at the electrolyte interface (i.e. for $x=x_{Stern}$). Only the solution 
with the correct asymptotic behavior has been selected, ensuring that $\phi\left(\left|x\right|\rightarrow\infty\right)=0$.
This model effectively 
corresponds to the Gouy-Chapman (GC) model for the diffuse layer\cite{Gouy1910, Chapman1913}, and shares the assumption 
of a planar distribution of charge at a fixed distance from the slab surface with the Helmholtz model described in Section \ref{subsub:helmholtz}. 
However, it includes a more physically sound shape of the diffuse layer along the surface normal.
In the linearized regime, the one-dimensional solution of the electrostatic problem would instead be given by
(see Supplemental material for the derivation):
\begin{equation}
\phi\left(x\right)=c_0\exp\left(\frac{k}{\varepsilon_0}\left|x\right|\right)\label{eq:LGCS_potential}
\end{equation}
where $c_0$ can be obtained by imposing continuity of the normal component of the electric field at the electrolyte interface.

For complex interfaces and general geometries, and for applications for which the 
linear-regime of the PBE is not expected to hold,
one needs to numerically solve the full non-linear PBE (Eq. \ref{eq:PBE}) to 
find the electrolyte concentration that minimizes the energy of the solvated system.

\subsubsection{Size-Modified Poisson-Boltzmann Model\label{subsub:MPB}}

The standard PB model assumes point-like  
ions, and consistently overestimates the 
electrolyte countercharge accumulation
at electrode surfaces. 
An improved model for the diffuse layer accounts for the steric repulsion 
between the ions, which opposes the electrostatic attraction 
towards the electrode surface and therefore limits electrolyte crowding. 
This is the so-called size-modified PB (MPB) model,
which can be derived from the free-energy functional as in Eq. \ref{eq:free-energy-PB}
but exploiting an entropy density expression that accounts for 
the finite-size of the ionic particles.

Borukhov et al. \cite{Borukhov-PRL-1997, Borukhov-ElectrochemActa-2000}
derived such an entropy expression from a lattice-gas model.
In particular, the volume of the continuum solution
is divided into a three dimensional lattice, with each cell of the lattice being 
occupied by no more than one ion. Thus, the cell volume $a^3$ or, equivalently, the 
maximum local ionic concentration $c_{0}=\frac{1}{a^3}$, sets the
distance of closest approach between ionic particles in the solution.
In this framework, the solute region is not part of the 
continuum solution and should therefore give zero contribution
to the solution entropy density. 
Otani and Sugino, who focused on two-dimensional slab systems, 
naturally achieved this limit by setting the boundary for the continuum solution region 
at a fixed distance from the surface\cite{otani_first-principles_2006}. 
In their derivations, Jinnouchi and Anderson \cite{Jinnouchi-PRB-2008} have instead imposed 
such a limit through an effective repulsive interaction 
between solute and electrolyte, which prevents the electrolyte solution
from entering the quantum-mechanical region.
Ringe et al. \cite{Ringe2016Function-Space-BasedDFT, Ringe2017TransferableSolutions}
have similarly accounted for such repulsive potential by recasting it in the form of an exclusion function. 
Here we follow a different approach\cite{Dabo-arXiv-2008, DaboThesis}, and enforce the limit by imposing a space-dependence
for the maximum ionic concentration, consequently exploiting the 
complementary interface function $\gamma\left(\mathbf{r}\right)$: 
$c_{0}\equiv c_{0} \left(\mathbf{r}\right)= c_{max}\gamma\left(\mathbf{r}\right)$.
The final expression for the electrolyte entropy density is therefore:
\begin{multline}
s\left[\left\{ c_{i}\left(\mathbf{r}\right)\right\} \right]=-k_{B}\sum_{i=1}^{p}c_{i}\left(\mathbf{r}\right)\ln\frac{c_{i}\left(\mathbf{r}\right)}{c_{max}\gamma\left(\mathbf{r}\right)}+\\
-k_{B}\left(c_{max}\gamma\left(\mathbf{r}\right)-\sum_{i=1}^{p}c_{i}\left(\mathbf{r}\right)\right)\ln\left(1-\sum_{i=1}^{p}\frac{c_{i}\left(\mathbf{r}\right)}{c_{max}\gamma\left(\mathbf{r}\right)}\right), \label{eq:entropy-MPB}
\end{multline}
The first and second terms in Eq. \ref{eq:entropy-MPB} can be identified as the 
entropy contributions from the ions and the solvent, respectively.
As in Eq. \ref{eq:entropy-PB}, the exclusion function $\gamma\left(\mathbf{r}\right)$ 
 sets the boundary of the region that contributes to the entropy of the electrolyte solution.

%As the standard PB model, also the MPB model can be alternatively derived 
%by assuming the solute-occupied volume as part of the electrolyte solution while simultaneously
%accounting for a repulsive interaction term that mimics the short-range interaction 
%between the solute and the ions. We note, however, that while the
%two approaches lead to identical derivations in the standard PB model, this is not the case for the size-modified 
%PB model, as further discussed in Appendix \ref{ap:alternative-formulation}. 

By minimizing the free-energy functional in Eq.
\ref{eq:free-energy-PB} with respect to the ion concentration we 
obtain the following expressions for the electrolyte chemical potentials (cf. Eq. \ref{eq:chemical-potential-PB}):
\begin{align}
\mu_{i}=k_{B}T\ln\left(\frac{c_{i}^{0}}{c_{max}-\sum_{i=1}^{p}c_{i}^{0}}\right),\label{eq:chemical-potential-MPB}
\end{align}
and for the equilibrium ionic concentrations (cf. Eq. \ref{eq:concentration-PB}):
\begin{align}
c_{i}\left(\mathbf{r}\right)=\frac{\gamma\left(\mathbf{r}\right)c_{i}^{0}e^{-\frac{z_{i}\phi\left(\mathbf{r}\right)}{k_{B}T}}}{1-\sum_{i=1}^{p}\frac{c_{i}^{0}}{c_{max}}\left(1-e^{-\frac{z_{i}\phi\left(\mathbf{r}\right)}{k_{B}T}}\right)} \equiv c^{MPB}_{i}\left(\phi\left(\mathbf{r}\right)\right) \label{eq:concentration-MPB}
\end{align}
The denominator in Eq. \ref{eq:concentration-MPB} 
renormalizes the concentration in the regions 
where the electrostatic interaction energy is comparable to or larger than $k_BT$, and sets $c_{max}$
as the maximum electrolyte concentration. This parameter can be related to the effective ionic
radius $r_i$ through $c_{max} = \frac{3P}{4\pi r_i^3}$. In the following, we will
assume random close packing for the electrolyte particles and correspondingly set the packing efficiency $P = 0.64$. 
Note that the point charge limit of Eq. \ref{eq:concentration-MPB}, which corresponds to
 $c_{max}\rightarrow \infty$, consistently leads to the equilibrium concentration 
 as derived in the standard PB model (Eq. \ref{eq:concentration-PB}).

By substituting Eqs. \ref{eq:entropy-MPB}-\ref{eq:concentration-MPB} into Eq. \ref{eq:free-energy-PB}, one obtains the following expression for the MPB free-energy functional: 
%\begin{multline}
%F^{MPB}\left[\rho\left(\mathbf{r}\right),\phi\left(\mathbf{r}\right)\right] =\int\left[-\frac{\epsilon\left(\mathbf{r}\right)}{8\pi}\left|\nabla\phi\left(\mathbf{r}\right)\right|^{2}+\rho\left(\mathbf{r}\right)\phi\left(\mathbf{r}\right)+ k_{B}Tc_{max}\gamma\left(\mathbf{r}\right)\ln\left(c_{max}-\sum_{i=1}^{p}c_{i}^{0}\right) + \right.\\
%\left. - k_BTc_{max}\gamma\left(\textbf{r}\right)\ln\left(c_{max}-\sum_{i=1}^{p}c_{i}^{0}\left(1-e^{-\frac{z_i \phi\left(\mathbf{r}\right)}{k_BT}}\right)\right) \right]\mathrm{d}\mathbf{r}+\\
% - k_{B}Tc_{max}V\ln\left(\frac{c_{max}-\sum_{i=1}^{p}c_{i}^{0}}{c_{max}}\right), \label{eq:freeenergyfunctional2}
%\end{multline}
\begin{multline}
F^{MPB}\left[\rho\left(\mathbf{r}\right),\phi\left(\mathbf{r}\right)\right] =\int\left[-\frac{\epsilon\left(\mathbf{r}\right)}{8\pi}\left|\nabla\phi\left(\mathbf{r}\right)\right|^{2}+\rho\left(\mathbf{r}\right)\phi\left(\mathbf{r}\right)+ k_{B}Tc_{max}\gamma\left(\mathbf{r}\right)\ln\left(c_{max}-\sum_{i=1}^{p}c_{i}^{0}\right) + \right.\\
\left. - k_BTc_{max}\gamma\left(\textbf{r}\right)\ln\left(c_{max}-\sum_{i=1}^{p}c_{i}^{0}\left(1-e^{-\frac{z_i \phi\left(\mathbf{r}\right)}{k_BT}}\right)\right) \right]\mathrm{d}\mathbf{r}
 - k_{B}Tc_{max}V\ln\left(\frac{c_{max}-\sum_{i=1}^{p}c_{i}^{0}}{c_{max}}\right), \label{eq:freeenergyfunctional2}
\end{multline}
where $V$ is the simulation cell volume. 
Minimization with respect to $\phi\left(\mathbf{r}\right)$ finally leads to the size-modified Poisson-Boltzmann equation (MPBE), which is analogous to 
the standard PBE (Eq. \ref{eq:PBE}) where, however, $c^{PB}_{i}\left(\phi\left(\mathbf{r}\right)\right)$ is replaced by $c^{MPB}_{i}\left(\phi\left(\mathbf{r}\right)\right)$.

\subsubsection{Additional Interactions\label{subsub:additional-interactions}}

The MPB model accounts for the steric repulsion between the ions  
in the solution, which limits electrolyte crowding.
In addition, the solute and the ionic particles are expected to be surrounded
by a solvation shell, where diffusing electrolyte particles are not expected 
to enter. 
The presence of this solvent-accessible but ion-free region, generally known as the Stern layer\cite{Stern1924},
can be simulated in a continuum framework via finite spacing between the onset 
of the dielectric function and the electrolyte charge density.

Such a spacing can be effectively introduced through 
 an \emph{ad-hoc} repulsive term 
between solute and electrolyte, $\varphi\left(\textbf{r}\right)$\cite{Jinnouchi-PRB-2008}.
The repulsive interaction would therefore give rise to the following free-energy contribution:
\begin{equation}
E_{rep}\left[\left\{ c_{i}\left(\mathbf{r}\right)\right\} \right] = \sum_{i=1}^{p}\int c_{i}\left(\mathbf{r}\right) \varphi\left(\mathbf{r}\right)\mathrm{d}\mathbf{r}, \label{eq:repulsive-energy}
\end{equation}
and subsequently appear in the expression for the equilibrium electrolyte concentration:
\begin{align}
c_{i}\left(\mathbf{r}\right)=\frac{\gamma\left(\mathbf{r}\right)c_{i}^{0}e^{-\frac{z_{i}\phi\left(\mathbf{r}\right) +\varphi\left(\textbf{r}\right)}{k_{B}T}}}{1-\sum_{i=1}^{p}\frac{c_{i}^{0}}{c_{max}}\left(1-e^{-\frac{z_{i}\phi\left(\mathbf{r}\right)+\varphi\left(\textbf{r}\right)}{k_{B}T}}\right)} \label{eq:concentration-SMPB}
\end{align}

As noted by Ringe et al.\cite{Ringe2016Function-Space-BasedDFT}, a repulsive solute-electrolyte interaction
 can be recast in the form of an electrolyte-specific exclusion function 
$\alpha\left(\textbf{r}\right)\equiv e^{-\frac{\varphi\left(\textbf{r}\right)}{k_bT}}$.
The exclusion function alone prevents the electrolyte from approaching and entering the solute region
in their Stern-corrected MPB model.  
The exclusion function $\gamma\left(\textbf{r}\right)$ that appears in our model has a 
different physical origin, since it reflects the hard separation between quantum-mechanical and continuum regions. 
While the presence of the Stern layer could be effectively included in our
model by using separate interface functions for dielectric and electrolyte,
as for instance done by Dabo et al.\cite{Dabo-arXiv-2008, DaboThesis},  
we find the picture of a single interface setting the boundary 
between quantum solute and continuum solution more physically sound.
We thus resort to repulsive interactions to introduce the finite spacing between the 
onsets of the dielectric and the electrolyte fluids. Note that, similarly to Ringe's model, our approach also predicts a zero entropic
contribution from the Stern-layer volume (cf. Eq. \ref{eq:entropy-MPB}), which is consistent with the expected
absence of diffusing solvent and electrolyte particles in this region.

For a two-dimensional system like a metal slab, we find appropriate to use 
one-dimensional exponential functions to define the repulsion potential: 
\begin{equation}
\varphi\left(x\right) = e^{-\frac{\left| x-x_0\right| - d}{w}}, \label{eq:repulsive-potential}
\end{equation}
where $x_0$ corresponds the $x$ coordinate of the slab center and the parameters
$d$ and $w$ set the position and decay rate of the potential, respectively. 

Baskin and Prendergast \cite{Baskin-JElectrochemSoc-2017} have proposed a similar formalism 
 to account for the specific adsorption of electrolyte species. 
 In particular, they have used a Morse-like potential in a fully-continuum 
 model to mimic anion adsorption on the electrode surface: 
\begin{equation}
\varphi\left(x\right) = E_{ads}\left(\left(1- e^{-\frac{\left| x-x_0\right| - d}{w}}\right)^2 - 1\right), \label{eq:attractive-potential}
\end{equation}
where $E_{ads}$ is the anion adsorption energy and $d$ now defines the 
 distance between the surface plane and the adsorbed anion species.
We have tested the introduction of such an interaction term in our mixed 
first-principles-continuum model. This description is computationally attractive as
it bypasses the need for surface configuration and
adsorbate coverage samplings. However, it is clear that such a model cannot be 
expected to capture electronic-structure changes of the metal surface 
beyond mean-field electrostatic effects.

\subsection{Computational Implementations\label{sub:numericalsolution}}

\subsubsection{Planar Countercharge Model\label{subsub:helmholtz-implementation}}

For all the models presented here, calculations are performed in a 
symmetric setup: the electrode 
surface is modeled by means of a two-dimensional slab exposing two identical
faces to the continuum solution. The computational setup thus involves 
two metal-solvent interfaces. Two charge distributions 
are added in front of the outermost atomic layers to compensate 
for the net charge of the surface, $q$.  

For numerical reasons, the sharp countercharge plane
that characterizes the Helmholtz model for the diffuse layer
is broadened to have a Gaussian-shaped profile along the surface
normal direction $x$:
 \begin{equation}
\rho_{ions}^{Helmholtz}\left(x\right) = \frac{q}{2A\sqrt{\pi}\Delta}e^{-\frac{\left(\left| x-x_0\right| - d\right)^2}{\Delta^2}}, \label{eq:Helmholtz}
\end{equation}
where $A$ is the surface area and the factor 2 at the denominator of the prefactor
arises from the symmetric setup employed. 
The distance $d$ from the slab center $x_0$ and the spread parameter $\Delta$
constitute the only two parameters in the model. 

The countercharge distributions are straightforwardly added to the 
total charge of the system. The corresponding electrostatic 
potential does not require self-consistency, allowing for fast and stable
simulations.

\subsubsection{Analytic Planar-Averaged Poisson-Boltzmann Model\label{subsub:stern}}

In many ways, modeling the diffuse layer via the analytical one-dimensional solution to the Poisson-Boltzmann problem can be seen as a straightforward modification of the Helmholtz approach, in which the shape of the planar countercharges is no longer given by a Gaussian envelope of arbitrary spread, but rather obtained from a physically sound model. 
However, simply inserting the diffuse layer as a charge distribution in the simulation cell would incur significant numerical problems: the analytical results for the electrolyte concentrations can present very sharp features close to the interface, which cannot be described accurately with the standard numerical resolution of the electronic-structure calculation.
Additionally, diffuse layers may have very long decaying length-scales, extending for tens of nanometers from the electrochemical interface, thus requiring large simulation cells. For these reasons, when possible a description in terms of the effects of the diffuse layer on the quantum-mechanical system, i.e. its electrostatic potential, is preferred. 

The assumption behind the model is the one of two sharp interfaces at a fixed distance $x_{Stern}$ from the slab center $x_{0}$, above and below the slab: the quantum-mechanical system is fully contained within the two interfaces, while the diffuse layer is fully in the outer regions and is uniform along the $yz$ planes perpendicular to the slab normal. With this setup, the net effect of the diffuse layer on the system would be a uniform shift, $\Delta\phi^{DL}$, of the electrostatic potential in the quantum-mechanical region of space, provided that the latter is computed with open-boundary conditions (OBC) along the $x$ axis. Thus, a possible definition of the full potential in the simulation cell is represented by
\begin{equation}
\phi\left(\mathbf{r}\right)=\begin{cases}
\phi^{OBC}\left(\mathbf{r}\right)+\Delta\phi^{DL}_0 & \left|x-x_0\right|<x_{Stern}\\
\phi^{DL}\left(x\right) & \left|x-x_0\right|\ge x_{Stern}\label{eq:correction_1}
\end{cases},
\end{equation}
where $\phi^{DL}$ is given by Eqs. (\ref{eq:GCS_potential}) or (\ref{eq:LGCS_potential}), and the shift is computed as
\begin{equation}
\Delta\phi^{DL}_0=\phi^{DL}\left(x_{Stern}\right)-\left\langle \phi^{OBC}\right\rangle _{yz}\left(x_{Stern}\right).
\end{equation}
Since the quantum-mechanical system is not bound to be perfectly homogeneous in the $yz$ plane, its planar average, $\left\langle\phi^{OBC}\right\rangle _{yz}\left(x\right)=A^{-1}\int\int\phi^{OBC}\left(\mathbf{r}\right)\mathrm{d}y\mathrm{d}z$, is used in the above equation, where $A$ is the surface area in the simulation cell. As a consequence, discontinuities can be present in the potential of Eq. (\ref{eq:correction_1}) when passing through the $x_{Stern}$ interfaces. Even though these discontinuities happen in a region of space which is not occupied by the quantum-mechanical system, they may be a source of numerical instabilities. To overcome this limitation, a slightly different path can be followed, where the diffuse-layer contribution to the potential is expressed as a one-dimensional continuous and smooth correction defined in the whole simulation cell, namely 
\begin{equation}
\phi\left(\mathbf{r}\right)=\phi^{OBC}\left(\mathbf{r}\right)+\Delta\phi^{DL}\left(x\right)\label{eq:correction_2}
\end{equation} where
\begin{equation}
\Delta\phi^{DL}\left(x\right)= \begin{cases} 
\Delta\phi^{DL}_0 & \left|x-x_0\right| < x_{Stern} \\ 
\phi^{DL}\left(x\right)-\langle \phi^{OBC}\rangle_{yz}\left(x\right) & \left|x-x_0\right| \ge x_{Stern}\end{cases}.
\end{equation}
The planar average of the potential on the right-hand side of the above equation can then be approximated by the one-dimensional potential of a planar-averaged charge distribution, namely
%\begin{multline}
%\langle\phi^{OBC}\rangle_{yz}\left(\left|x-x_0\right|>x_{Stern}\right)\approx\\
%\langle\phi^{OBC}\rangle_{yz}\left(x_{Stern}\right)-\frac{2\pi q}{A \epsilon_0}\left|x-x_{Stern}\right|,
%\end{multline} 
\begin{equation}
\langle\phi^{OBC}\rangle_{yz}\left(\left|x-x_0\right|>x_{Stern}\right)\approx\langle\phi^{OBC}\rangle_{yz}\left(x_{Stern}\right)-\frac{2\pi q}{A \epsilon_0}\left|x-x_{Stern}\right|,
\end{equation} 
where $q$ is the total charge of the quantum-mechanical system and we have used the classical electrostatics result for the potential of a planar charge distribution. With the above formulation, the correction and its first derivative are both continuous at the interface. 
 
While the correction described here is similar in spirit to the 
`electrochemical boundary conditions' from Refs. \cite{DaboThesis, Dabo-arXiv-2008, Dabo2010},
our approach makes use of the electrostatic potential that analytically solves the 
PBE in order to determine the diffuse-layer contribution to $\phi\left(\textbf{r}\right)$, 
without the need of an iterative procedure. Another novel element of the procedure described
is the implementation of the correction that corresponds to the linear-regime version of the PB problem, 
which allows for the validation of the corresponding numerical approach.

While the above correction is defined for open-boundary conditions, standard electronic-structure simulations usually exploit periodic-boundary conditions. In this case, an alternative expression of the electrostatic potential of the electrochemical interface can be obtained, which incorporates the handling of PBC artifacts and of the diffuse layer into a single continuous and smooth one-dimensional correction. In particular, an approximate OBC potential is obtained as
\begin{equation}
\phi^{OBC}\left(\mathbf{r}\right)\approx\phi^{PBC}\left(\mathbf{r}\right)+\Delta\phi^{2D}\left(x\right) \label{eq:parabolic}
\end{equation} and assuming a planar-averaged charge distribution a parabolic correction can be expressed as \cite{Andreussi-PRB-2014}
\begin{equation}
\Delta\phi^{2D}\left(x\right)=\frac{\alpha_{1D}}{L_{x}}q-\frac{2\pi q}{V}x^{2}+\frac{4\pi}{V}d_{x}x-\frac{2\pi}{V}Q_{xx}, \label{eq:parabolic-2}
\end{equation}
where $\alpha_{1D}=\pi/3$ is the Madelung constant of a one-dimensional lattice, $V$ is the cell volume, $q$, $d_x$, and $Q_{xx}$ are the monopole, dipole, and quadrupole moments along the $x$ axis of the charge distribution. The corrected electrostatic potential can thus be easily expressed in terms of the PBC one as 
\begin{equation}
\phi\left(\mathbf{r}\right)\approx\phi^{PBC}\left(\mathbf{r}\right)+\Delta\phi^{2D}\left(x\right)+\Delta\phi^{DL}\left(x\right).
\end{equation}
The above approach can be adopted for any situation where an analytical one-dimensional solution to the electrostatic problem is available. In the description of the diffuse layer, both the Gouy-Chapman model and its linearized version have analytical solutions that can be inserted into the $\phi^{DL}\left(x\right)$ term. Although based on a planar-average approximation, this class of correction approaches has significant advantages, in terms of speed and stability, when compared to more advanced numerical solutions of the electrostatic equations.

\subsubsection{Linearized Poisson-Boltzmann Model\label{subsub:linearized-pb}}

In order to tackle the linear-regime version of the PB problem, 
we solve the corresponding differential equation using a preconditioned gradient-based method 
as proposed by Fisicaro et al. \cite{Fisicaro2016a}, which we will only summarize here. 
Briefly, 
we apply a conjugate-gradient algorithm to solve the LPBE:
\begin{equation}
\underbrace{\left(\nabla\cdot\varepsilon\left(\mathbf{r}\right)\nabla-k^2\gamma\left(\mathbf{r}\right)\right)}_{\mathbf{A}}\phi\left(\mathbf{r}\right)=\underbrace{-4\pi\rho\left(\mathbf{r}\right)}_{b\left(\mathbf{r}\right)}, \label{eq:LMPBEalgo}
\end{equation}
using the following preconditioning operator:
\begin{equation}
\mathbf{P} = \sqrt{\varepsilon\left(\mathbf{r}\right)} \nabla^2 \sqrt{\varepsilon\left(\mathbf{r}\right)}
\end{equation}
Therefore, instead of minimizing the 
residual function $r\left(\mathbf{r}\right) = b\left(\mathbf{r}\right) - \mathbf{A}\phi\left(\mathbf{r}\right) $, 
one finds the solution of the problem by minimizing the 
preconditioned residual $v\left(\mathbf{r}\right) = \mathbf{P}^{-1}r\left(\mathbf{r}\right) =  \mathbf{P}^{-1}\left(b\left(\mathbf{r}\right) - \mathbf{A}\phi\left(\mathbf{r}\right) \right)$.
The algorithm has been proven to converge 
 in a limited number of iterations for simple analytical cases\cite{Fisicaro2016a}. 
In addition, the choice of the preconditioner minimizes the computational effort 
required \cite{Fisicaro2016a}. 
In fact, the action of the operator $\mathbf{A}$ on the preconditioned residual $v_n\left(\mathbf{r}\right)$, 
which needs to be computed at each of the $n$-th solver's iteration, can be efficiently 
estimated as:
\begin{align}
\begin{split}
\mathbf{A} v_n\left(\mathbf{r}\right) &= \left(\nabla\cdot\varepsilon\left(\mathbf{r}\right)\nabla-k^2\gamma\left(\mathbf{r}\right) \right) v_n\left(\mathbf{r}\right)\\
&=\left(\varepsilon\left(\mathbf{r}\right)\nabla^2 +\nabla\varepsilon\left(\mathbf{r}\right)\cdot\nabla -k^2\gamma\left(\mathbf{r}\right) \right) v_n\left(\mathbf{r}\right) \\
&= -\left( q\left(\mathbf{r}\right) + k^2\gamma\left(\mathbf{r}\right) \right)v_n\left(\mathbf{r}\right) + r_n\left(\mathbf{r}\right),
\end{split}
\end{align}
where $q\left(\mathbf{r}\right) = \sqrt{\varepsilon\left(\mathbf{r}\right)} \nabla^2 \sqrt{\varepsilon\left(\mathbf{r}\right)}$ and 
we have used $r_n\left(\mathbf{r}\right) = \mathbf{P}v_n\left(\mathbf{r}\right) = \varepsilon\left(\mathbf{r}\right)\nabla^2v_n\left(\mathbf{r}\right)  +\nabla\varepsilon\left(\mathbf{r}\right)\cdot\nabla v_n\left(\mathbf{r}\right) + q\left(\mathbf{r}\right) v_n\left(\mathbf{r}\right)$.
The term $q\left(\mathbf{r}\right)$ can be evaluated only once and stored in memory. 
Once the terms $r_n\left(\mathbf{r}\right)$ and $v_n\left(\mathbf{r}\right) $ are computed, the 
evaluation of $\mathbf{A} v_n\left(\mathbf{r}\right)$ at each of the following iterations 
 requires only vector-vector multiplications. 
The bottleneck of the algorithm is represented 
by the calculation of $v_n\left(\mathbf{r}\right)$, which is calculated as\cite{Fisicaro2016a}:
\begin{equation}
v_n\left(\mathbf{r}\right) = \mathbf{P}^{-1} r_n\left(\mathbf{r}\right) = \frac{1}{\sqrt{\varepsilon\left(\mathbf{r}\right)}}\left(\nabla^2\right)^{-1}\left( \frac{r_n\left(\mathbf{r}\right)}{\sqrt{\varepsilon\left(\mathbf{r}\right)}} \right).
\end{equation}
The overall algorithm performance is thus highly dependent on the solution of the 
standard Poisson problem, which in our case is carried out in reciprocal space. 
The term
$r_n\left(\mathbf{r}\right)$ is computed instead from the knowledge of the residual function 
at the previous step $r_{n-1}\left(\mathbf{r}\right)$ and other quantities derived from
the preconditioned residual $v_n\left(\mathbf{r}\right)$\cite{Fisicaro2016a}, as typically carried 
out in conjugate-gradient approaches.

While the algorithm from Fisicaro \emph{et al.} was tested in Ref.\cite{Fisicaro2016a} by solving the LPBE only
for simple analytic potentials, we investigate here for the first time the 
performance of such preconditioned conjugate-gradient algorithm for realistic
electrified interfaces through our novel implementation in the ENVIRON module\cite{ENVIRON} 
for Quantum ESPRESSO\cite{Giannozzi2009QMaterials, Giannozzi2017}.

\subsubsection{Standard and Size-Modified Poisson-Boltzmann Model\label{subsub:full-pb}}

The preconditioned conjugate gradient algorithm from Ref. \cite{Fisicaro2016a} can
only tackle linear problems, like the one represented by the 
linearized-PB equation.
Fisicaro \emph{et al.} have also proposed an iterative algorithm devised to solve the full non-linear PB equation\cite{Fisicaro2016a}, which, however,
turned out not to be sufficiently stable to deal with extended charged systems.
For the numerical solution of the full (size-modified) PB equation,
we thus resort to the more robust Newton-based algorithm as proposed by Ringe et al. \cite{Ringe2016Function-Space-BasedDFT},
which we have also implemented in the development version of the ENVIRON module\cite{ENVIRON}.
In particular, the free-energy functional minimization that leads to the 
PB equation is recast as a root-finding problem:
\begin{equation}
G\left[ \phi\left(\mathbf{r}\right)\right] = 0,  \label{eq:g}
\end{equation}
where $G\left[ \phi_{n}\left(\mathbf{r}\right) \right]$ is the $\phi\left(\mathbf{r}\right)$ functional derivative of
the free-energy $F\left[\phi\left(\mathbf{r}\right) \right]$:
%\begin{multline}
%G\left[ \phi_{n}\left(\mathbf{r}\right) \right] = \nabla\cdot\varepsilon\left(\mathbf{r}\right)\nabla\phi_n\left(\mathbf{r}\right)+4\pi\rho\left(\mathbf{r}\right)+\\
%+4\pi\sum_i^p z_i c_{i}\left(\phi_n\left(\mathbf{r}\right)\right), \label{eq:g2}
%\end{multline}
\begin{equation}
G\left[ \phi_{n}\left(\mathbf{r}\right) \right] = \nabla\cdot\varepsilon\left(\mathbf{r}\right)\nabla\phi_n\left(\mathbf{r}\right)+4\pi\rho\left(\mathbf{r}\right)+4\pi\sum_i^p z_i c_{i}\left(\phi_n\left(\mathbf{r}\right)\right), \label{eq:g2}
\end{equation}
Following Newton's iterative algorithm, the estimate for $\phi\left(\mathbf{r}\right)$ at the the $n$-th step is obtained as:
\begin{equation}
\phi_{n+1}\left(\mathbf{r}\right) = \phi_{n}\left(\mathbf{r}\right) + \frac{G\left[ \phi_{n}\left(\mathbf{r}\right) \right]}{G^{\prime}\left[ \phi_{n}\left(\mathbf{r}\right) \right]}. \label{eq:newton}
\end{equation}
Here $G^{\prime}\left[ \phi_{n}\left(\mathbf{r}\right) \right]$ is the Fr\'{e}chet derivative of $G\left[ \phi_{n}\left(\mathbf{r}\right) \right]$:
\begin{equation}
G^{\prime}\left[\phi_{n}\left(\mathbf{r}\right) \right]  = \nabla\cdot\varepsilon\left(\mathbf{r}\right)\nabla+4\pi\sum_i^p z_i\frac{\partial c_i}{\partial\phi}\left(\phi_n\left(\mathbf{r}\right)\right). \label{eq:g-prime}
\end{equation}
By rearranging the terms in Eq. \ref{eq:newton}, the following linear problem is recovered:
%\begin{multline}
%\left(\nabla\cdot\varepsilon\left(\mathbf{r}\right)\nabla+ 4\pi\sum_i^p z_i\frac{\partial c_i}{\partial\phi}\left(\phi_n\left(\mathbf{r}\right)\right) \right)\phi_{n+1}\left(\mathbf{r}\right) = \\-4\pi\left(\rho\left(\mathbf{r}\right) + \sum_i^p z_ic_i\left(\phi_n\left(\mathbf{r}\right)\right)+\right.\\ 
%\left.-\sum_i^p z_i\frac{\partial c_i}{\partial\phi}\left(\phi_n\left(\mathbf{r}\right)\right)\phi_n\left(\mathbf{r}\right) \right). \label{eq:newton-linear}
%\end{multline}
\begin{multline}
\left(\nabla\cdot\varepsilon\left(\mathbf{r}\right)\nabla+ 4\pi\sum_i^p z_i\frac{\partial c_i}{\partial\phi}\left(\phi_n\left(\mathbf{r}\right)\right) \right)\phi_{n+1}\left(\mathbf{r}\right) = \\-4\pi\left(\rho\left(\mathbf{r}\right) + \sum_i^p z_ic_i\left(\phi_n\left(\mathbf{r}\right)\right)-\sum_i^p z_i\frac{\partial c_i}{\partial\phi}\left(\phi_n\left(\mathbf{r}\right)\right)\phi_n\left(\mathbf{r}\right) \right). \label{eq:newton-linear}
\end{multline}

Overall, the algorithm proceeds as follows: starting from an initial guess for 
$\phi\left(\mathbf{r}\right)$, i.e. $\phi_{0}\left(\mathbf{r}\right)$, the charge
and screening terms in
Eq. \ref{eq:newton-linear} are evaluated. The new guess for the electrostatic potential, $\phi_{1}\left(\mathbf{r}\right)$,
is then determined by solving the corresponding linear problem using the preconditioned conjugate-gradient
procedure from Ref.  \cite{Fisicaro2016a}, as described in Section \ref{subsub:linearized-pb}. The charge and screening 
terms are updated, and the new linear problem solved to find $\phi_{2}\left(\mathbf{r}\right)$. These steps are
repeated until convergence is achieved.

\subsection{Computational Details\label{sub:computational-details}}

\begin{figure}%++++++++++++++++++++++++++++++++++++++++++++++++++++++
\begin{centering}
	\includegraphics[width=0.5\columnwidth]{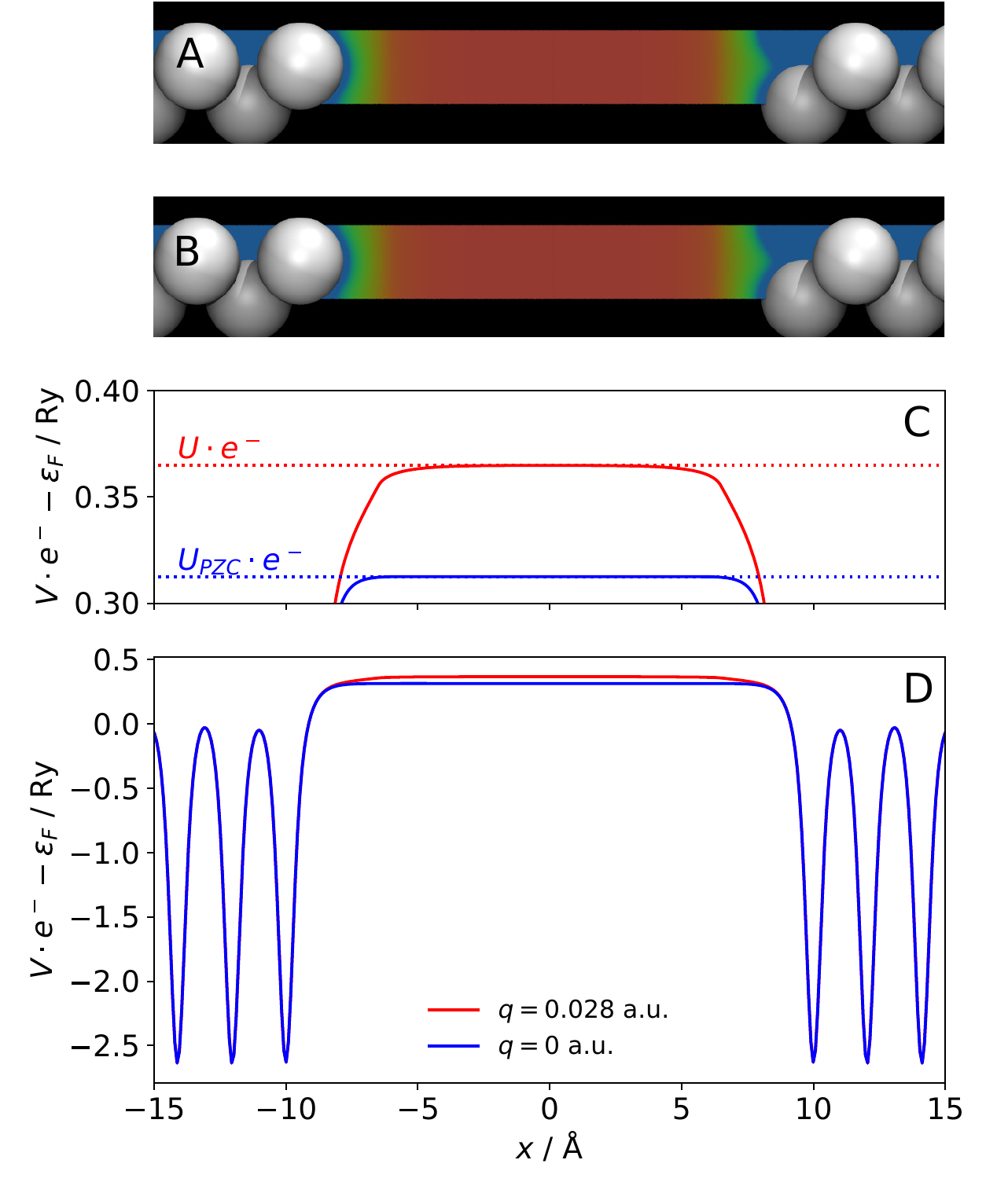}\\
\end{centering}
\caption{
%The Ag(100) slab employed in the calculation is sketched in the top 
%panel. Red and blue color maps illustrate the local values of the dielectric function $\varepsilon(\textbf{r})$ 
%and the local electrolyte charge density $\rho_{ions}(\textbf{r})$ as computed with the SSCS model with
%$c_0 = 10$ M and an overall slab charge $q = $ XX a.u.. In the bottom panel, 
2D-maps of the electrostatic potential computed with the planar-averaged analytical PB model (A) and the corresponding
numerical version (B) for a Ag(100) slab in vacuum with a total charge $q = $ 0.028 a.u.. 
The planar-averaged electrostatic potential computed for the neutral and the charged surface (red and blue, respectively)
is plotted as a function of the 
surface normal direction $x$ in C and D. The same data is plotted in the C and D panels, 
using different scales for the potential axis. The planar interface with a symmetric monovalent 
electrolyte with bulk concentration $c_0 = 0.1$ M has been employed.}
\label{fig:electrostatic}
\end{figure}

All the electrolyte models discussed have been implemented in the developer version 
of the ENVIRON module\cite{ENVIRON} for the Quantum ESPRESSO distribution\cite{Giannozzi2009QMaterials, Giannozzi2017}. 
Differential capacitances have been
calculated by numerically differentiating charge-potential curves. We have employed a 
canonical approach: we perform constant charge calculations and determine 
the applied potential $U$ \emph{a posteriori} from the difference between the asymptotic 
electrostatic potential and the Fermi energy of the system. Experimental data and simulations
have been compared to each other using as potential reference the corresponding 
estimate of the potential of zero charge. 
In the simulations, this is the potential computed for a neutrally-charged surface, 
$U_{PZC}$ (see Figure \ref{fig:electrostatic}).

%+++++++++++++++++++++++++++++++++++++++++++++++++++++++

All calculations have been performed 
using the Perdew-Burke-Ernzerhof exchange-correlation functional\cite{Perdew1996GeneralizedSimple, Perdew-PRL-1997} 
and pseudo-potentials from the GBRV set\cite{Garrity-ComputMatSci-2014}, 
which have been chosen according to guidelines from the Standard Solid-State Pseudopotential 
library\cite{Prandini-arXiv-2018} (SSSP efficiency 0.7). Cutoff energies for the plane wave and density expansions 
have been set to 35 Ry and 350 Ry, respectively. A $\Gamma$-centered 18x18x1 k-point grid 
(or equivalent) has been employed to sample the first Brillouin zone, using the cold smearing 
technique from Ref.\cite{marzarivanderbilt_prl_1999}
 with smearing parameter $\sigma = 0.01$ Ry. 

The Ag(100) slab has been constructed using 8 atomic layers at the 
bulk equilibrium lattice constant (4.149 \AA). The slab has been 
fully relaxed only in vacuum. While electrolyte- and solvent-related contributions to the atomic forces 
are accounted for and they in principle allow for fully self-consistent 
optimizations in the presence of the embedding continuum, test calculations for the present case show that further relaxations
 in implicit solvent negligibly affect the surface structure and the Fermi energy of the system. 

Particular care needs to be taken with respect to the simulation cell size when using the SSCS (`soft-sphere') cavity.
If the diameter of the atom-centered spheres exceeds one of the cell lattice vectors, a sharp 
transition arises in the region where the spheres overlap with their neighboring periodic replicas, which 
can trigger numerical instabilities. 
In order to obtain a smooth interface function, one has to set the simulation cell size such that all 
the sphere diameters fit the simulation box. For this reason, calculations using the SSCS have been 
performed in a (2x2) supercell. We also note that the Ag sphere radii as part of the original SSCS parameterization\cite{Fisicaro2017}
are such that the resulting cavity includes non-physical dielectric
pockets inside the metal slab. 
This issue has been fixed by introducing a non-local correction based on the the convolution of the interface function
with a solvent-size-related probe function. In this way, dielectric pockets that are smaller than the chosen
solvent molecule (in this case water) can be identified and removed. Further details on the 
construction of such non-local interface are deferred to a forthcoming publication \cite{Andreussi-JCTC-2018}.

The models characterized by a self-consistent optimization of the ionic countercharge density 
require large separations between periodic replicas of the slab along the surface normal. This is to account for 
the long-range electrolyte screening, whose typical length is the Debye length $\lambda_D$. Due to the partial
screening of the electrolyte charge by the dielectric, calculations that include implicit solvent require larger cell sizes
along the surface normal. We have verified that Fermi energies are converged within few meVs for a 20 \AA\ (60 \AA) 
separation between periodic images for calculations in vacuum (implicit solvent). 
For calculations involving particularly low bulk ionic concentrations ($\le 0.04$ M) 
we have doubled these spacings. Note that the planar-averaged implementation of the PB model does not require these large spacings,
as one resorts to the analytical solution of the one-dimensional problem to set the electrostatic potential at the cell boundaries. 
Such calculations have been thus performed with a spacing of 20 \AA\ between periodic images. 
 For both the numerical and analytic electrolyte models, we have made use of the parabolic
 corrective scheme from Ref. \cite{Andreussi-PRB-2014} in order to recover the potential of the isolated system
 from the electrostatic potential computed with periodic-boundary conditions 
 (see also Equations \ref{eq:parabolic} and \ref{eq:parabolic-2}). This correction guarantees that
 the electrostatic potential approaches zero at large distances from the metal slab, provided 
 that enough empty space is included in the unit cell for the numerical models.
 While charge neutrality is often enforced by means of a Lagrange multiplier $\mu_{el}$
  in the search for the electrostatic potential $\phi\left(\textbf{r}\right)$ that minimizes the energy of the system \cite{Tamashiro2003, Gunceler2013TheSystems, Melander2018},
the asymptotically-zero reference potential can be used with $\mu_{el}\equiv 0$.
This choice simultaneously provides the correct asymptotic limits for 
the electrolyte charge density and for the ionic concentration profiles\cite{Melander2018}.

%%%%%%%%%%%%%%%%%%%%%%%%%%%%%%%%%%%%%%%%%%%%%%%%%%%%%%%%%%%%%%%%%%%%%%
\section{RESULTS AND DISCUSSION\label{sec:Results-and-Discussion}}

\subsection{Vacuum\label{sub:results-vacuum}}

We start
by considering the differential capacitance (DC) of Ag(100) in a solution with the vacuum dielectric constant 
($\varepsilon_0 = 1$), which allows us to disentangle the electrolyte effects from the role played by the dielectric medium. 

\begin{figure}%++++++++++++++++++++++++++++++++++++++++++++++++++++++
\begin{centering}
	\includegraphics[width=0.5\columnwidth]{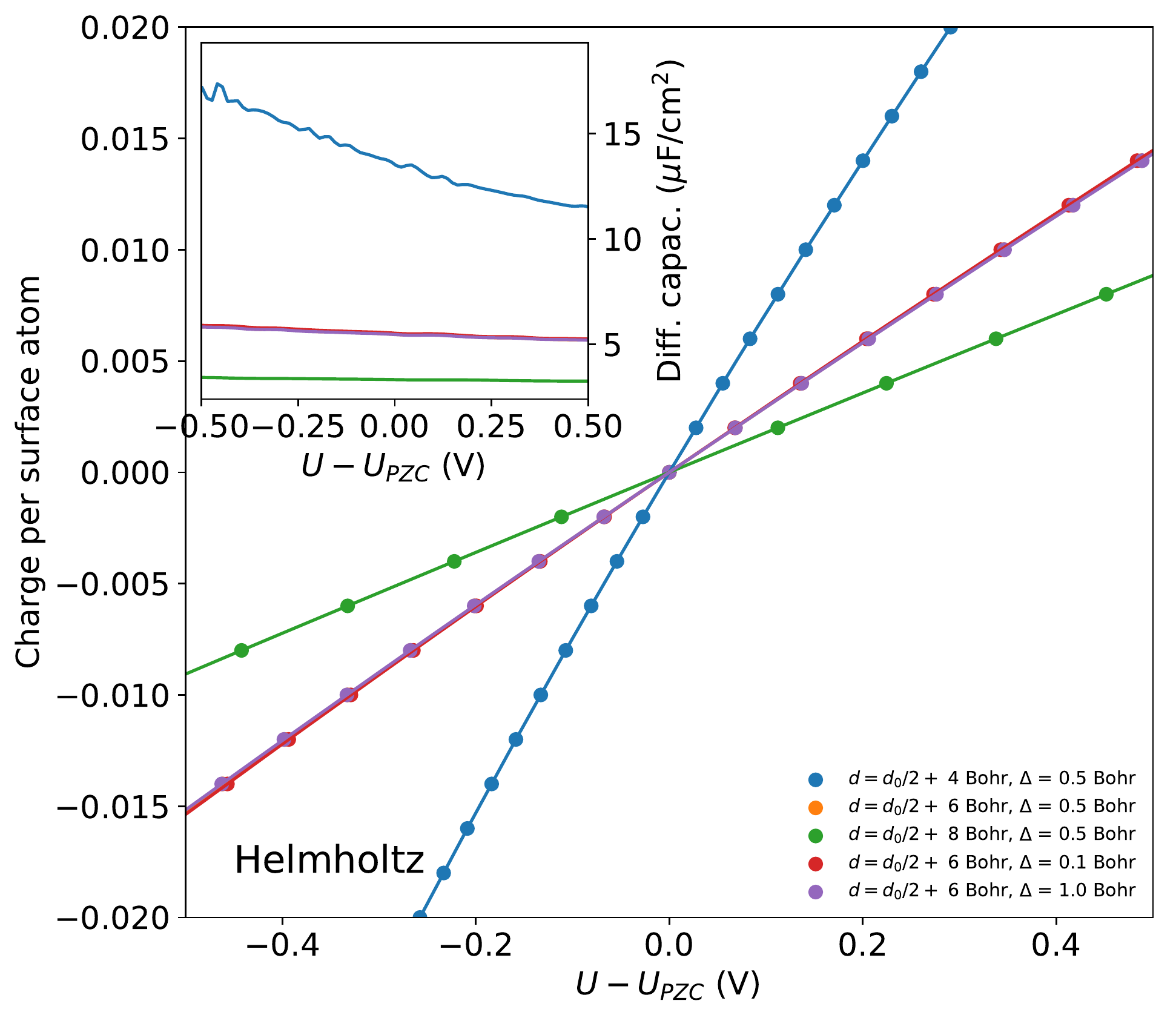}\\
\end{centering}
\caption{The charge per surface atom (in a.u.) is plotted as a function of the potential (in V). The 
Helmholtz model has been used for all data sets, varying the $d$ and $\Delta$ parameters that define 
the position and the width of the countercharge density, respectively. 
Note that the three lines corresponding 
to $d = d_0/2\ +\  6$ Bohr are essentially superimposed ($d_0$ is the slab thickness).  
The inset shows the differential capacitance as a function of the potential, 
as calculated from the analytical derivative of the spline-interpolated charge-potential
curves (same line styles as in the main plot). }
\label{fig:helmholtz-eps01}
\end{figure}

We first consider the planar-countercharge Helmholtz model (see Sections \ref{subsub:helmholtz} and \ref{subsub:helmholtz-implementation}), which represents the lowest-rung diffuse layer model among the ones considered here.
Figure \ref{fig:helmholtz-eps01} illustrates how the two parameters in the model, i.e. the surface-countercharge distance
$d$ and the spread of the charge distribution $\Delta$, affect the computed DC.
Overall, charge-potential curves are found to be close to linear for all tested parameter values.
The DC predicted by the Helmholtz model is thus almost potential independent,
with a small DC decrease for increasing potentials. This trend is consistent with
the larger electron-density spilling at more negative potentials, which effectively reduces 
the distance between the surface 
and the fixed countercharge distribution.
This simple capacitor model also explains the effect of the 
$d$ parameter on the computed DC,
as we observe
an increase (decrease) of the DC value 
for an inward (outward) shift of the neutralizing counterion density.
The broadening of the electrolyte charge density, as regulated by the $\Delta$ parameter, 
instead has a negligible effect on the DC. 
This is expected, as the spread of the distribution only affects the 
field in the narrow region where the countercharge is located. In contrast, the electrostatic 
potential at large distances from the countercharge planes is essentially unaffected by the $\Delta$
parameter, and so is the DC.
%We note, however, that the value of the spread parameter should be chosen such that the overlap between 
%the external charge and the surface electron density is negligible.

%+++++++++++++++++++++++++++++++++++++++++++++++++++++++

\begin{figure}%++++++++++++++++++++++++++++++++++++++++++++++++++++++
\begin{centering}
	\includegraphics[width=0.5\columnwidth]{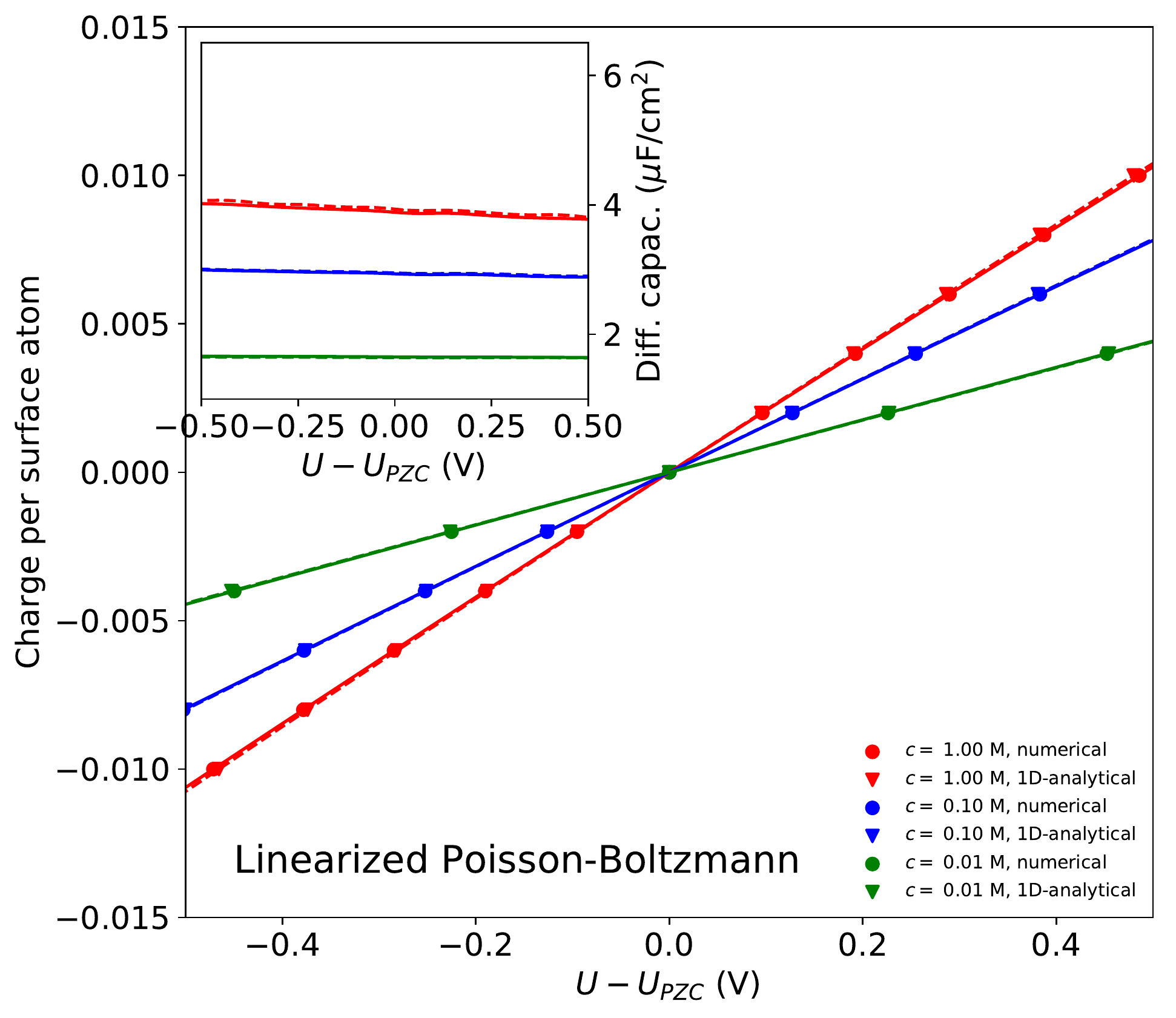}\\
\end{centering}
\caption{Same as Figure \ref{fig:helmholtz-eps01}, but for the planar-averaged analytic linearized-PB model 
(triangles and dashed lines) and the corresponding numerical implementation (circles and solid lines). 
Red, blue and green symbols correspond to bulk electrolyte concentrations $c_0 = 1$ M,
$c_0 = 0.1$ M and $c_0 = 0.01$ M, respectively.
$x_{Stern}$ in the analytic model (Section \ref{subsub:stern})
and $d$ in the planar interface employed in the numerical model (Section \ref{subsub:linearized-pb}) are set so 
that the interface lies in both cases 6.568 Bohr away from the surface.}
\label{fig:stern-linearized-eps01}
\end{figure}

As already mentioned in Section \ref{subsub:helmholtz}, the Helmholtz model for the diffuse layer 
does not include any dependence on the bulk electrolyte concentration. 
The planar-averaged PB model overcomes this limitation 
while retaining the assumption of a planar countercharge-density profile. 
Results obtained with the linearized-PB model (see Sections \ref{subsub:PB} and \ref{subsub:stern}) are
illustrated in Figure \ref{fig:stern-linearized-eps01}, which shows the computed 
charge-potential curves and corresponding capacitance values 
for three representative electrolyte concentrations.
The linear-regime PB model predicts a weak potential 
dependence of the DC, as also observed
for the Helmholtz model, but the computed capacitance now depends on the 
electrolyte concentration. In particular, lower DC values correspond to lower ionic 
concentrations. 

%+++++++++++++++++++++++++++++++++++++++++++++++++++++++

Figure \ref{fig:stern-linearized-eps01} also includes results of calculations performed with 
the numerical linearized PB solver (see Sections \ref{subsub:PB} and \ref{subsub:linearized-pb}),
using a planar but smooth interface function with an error-function profile along the surface normal.
We have used here a small spread parameter (0.01 Bohr)
in order to better compare results to the planar-averaged LPB model, in which 
a sharp planar interface defines the boundary of the region where the 
one-dimensional LPBE is analytically solved.
The DC computed through the numerical solution
of the LPBE agrees well with what obtained through the corresponding
planar-averaged analytic model. This is consistent with the interface
being essentially two-dimensional, as expected for closely-packed metal surfaces like Ag(100). 

\begin{figure}%++++++++++++++++++++++++++++++++++++++++++++++++++++++
\begin{centering}
	\includegraphics[width=0.5\columnwidth]{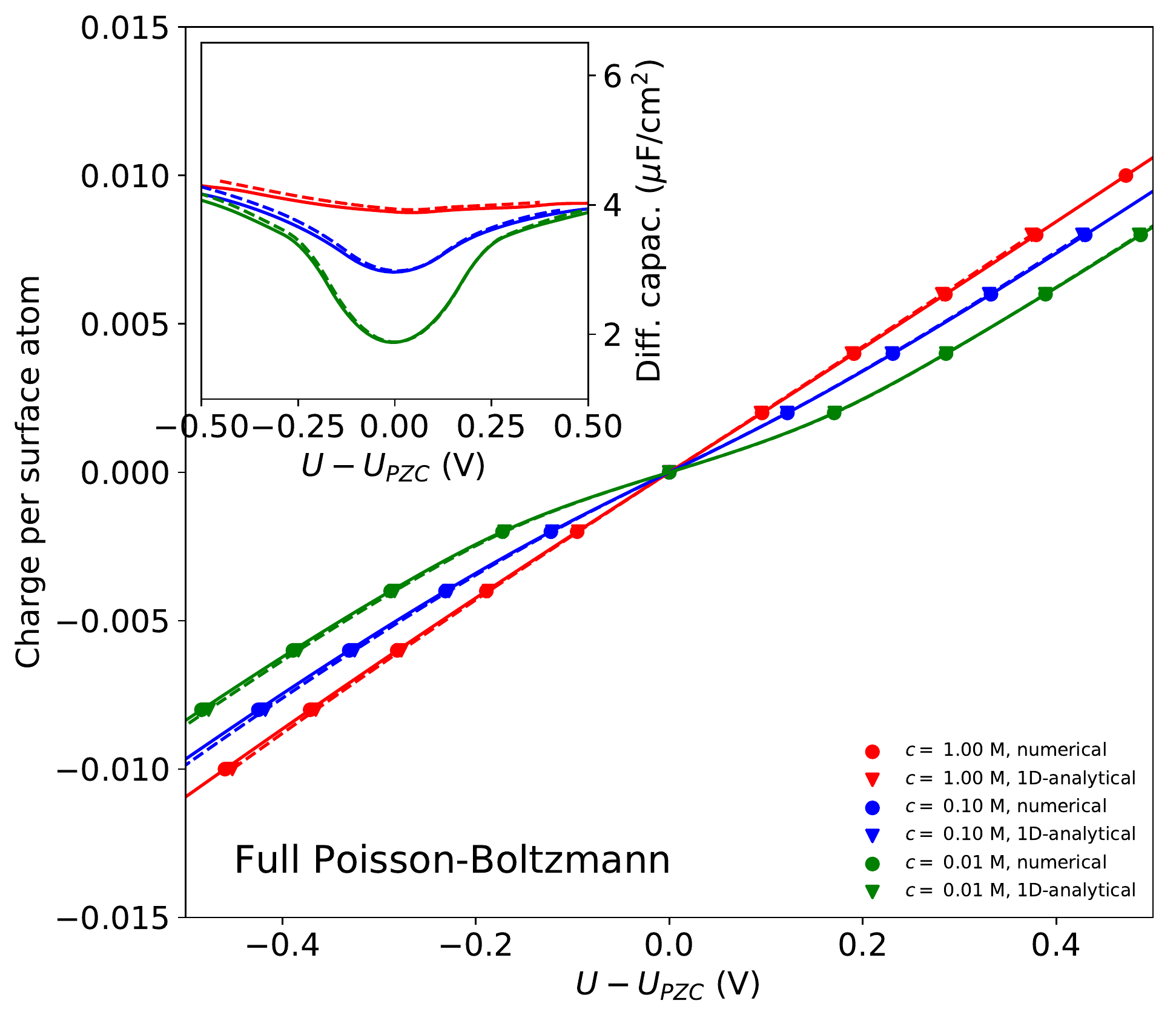}\\
\end{centering}
\caption{Same as Figure \ref{fig:stern-linearized-eps01}, but for the full PB model, in the 
planar-averaged analytic implementation (triangles and dashed lines) and 
the numerical implementation (circles and solid lines).}
\label{fig:stern-eps01}
\end{figure}%+++++++++++++++++++++++++++++++++++++++++++++++++++++++

The capacitance-potential trends obtained from the solution of the full-PBE (see Sections \ref{subsub:PB} and \ref{subsub:full-pb}) are quite different,
 as illustrated in Figure \ref{fig:stern-eps01}.  In contrast with the linear-regime model, 
 the DC curves computed with the non-linear electrolyte model exhibit a concentration-dependent drop at the potential 
 of zero charge (PZC), while similar capacitance values are observed at the largest potentials 
 simulated for all electrolyte concentrations (see also Figs. 4-10 and 4-11 of Ref.\cite{DaboThesis}). 
As also observed for the linearized model, we find very good agreement between 
the capacitance curves computed with the planar-averaged approach, which exploits 
the analytical solution of the PBE along the surface normal, and the full numerical 
implementation. As illustrated in Figure \ref{fig:electrostatic}, in fact, the electrostatic potential obtained with the 
numerical model is not significantly corrugated in the $yz$-plane at sufficiently large distance from the surface, 
and is thus very similar to the potential computed with the planar-averaged analytical approach.

\begin{figure}%++++++++++++++++++++++++++++++++++++++++++++++++++++++
\begin{centering}
	\includegraphics[width=0.5\columnwidth]{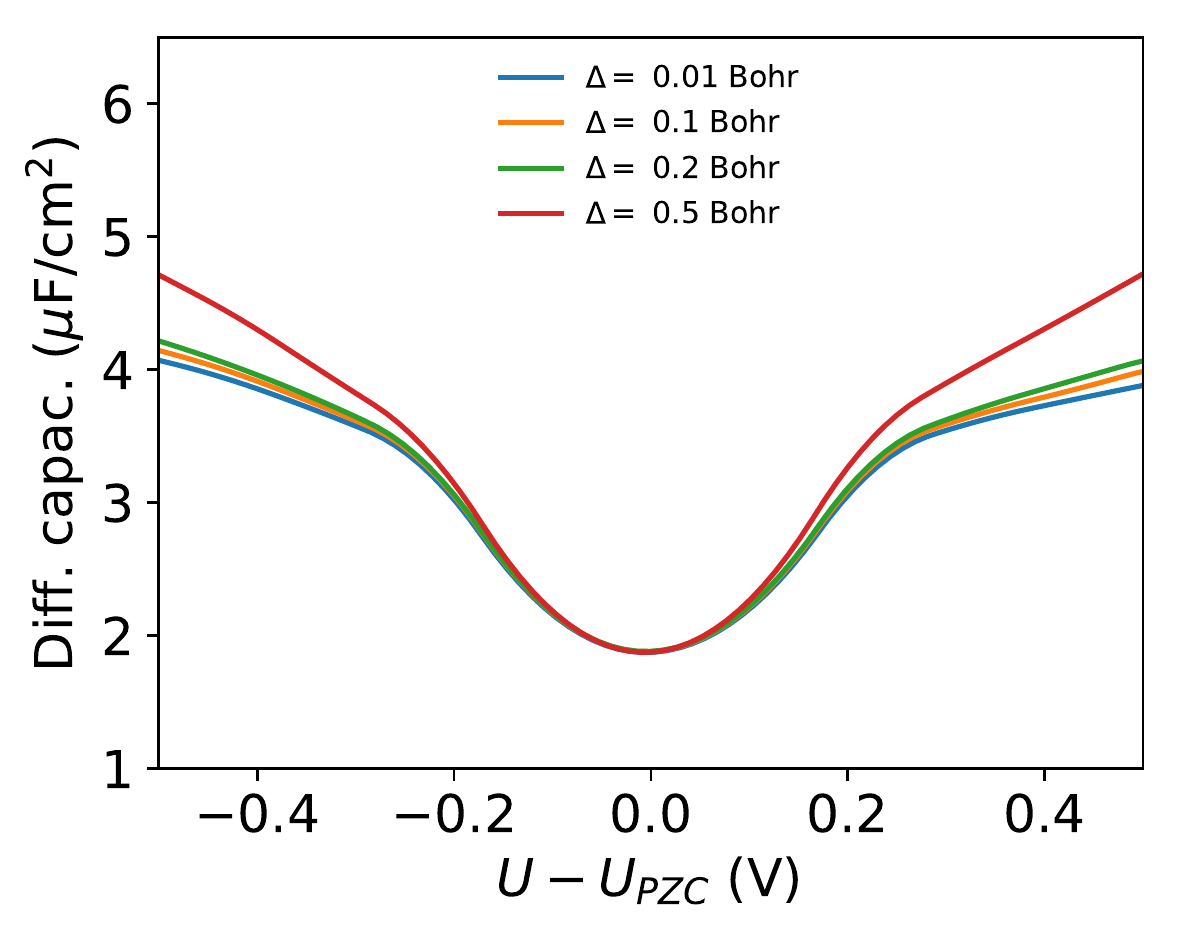}\\
\end{centering}
\caption{Full Poisson-Boltzmann model: differential capacitance as a function of the potential for various
values of the spread parameter $\Delta$ in the planar interface function, using the numerical solver. 
The distance parameter $d$
has been set so that the interface lies 6.568 Bohr away from the surface. The full numerical 
PB model has been used for all calculations, using $c_0 = 0.01$ M.}
\label{fig:sys-spread-eps01}
\end{figure}

The effect of the interface broadening on the DC
is illustrated in Figure \ref{fig:sys-spread-eps01}, where we compare DC-potential curves computed with the 
numerical solver of the full-PBE (Section \ref{subsub:full-pb}) and a
planar but smooth interface function. We have tested different values of the spread parameter $\Delta$, 
ranging from 0.01 to 0.5 Bohr. 
The DC is found to increase for increasing values of $\Delta$, which follows from the onset 
of the electrolyte-accessible region becoming closer to the surface. 
This effect is most pronounced at large (absolute) potentials and at high electrolyte concentrations.
Under such conditions, in fact, the electrolyte charge density at the interface boundary is larger 
and sharper, and thus more sensitive to small changes in the onset region. 

%+++++++++++++++++++++++++++++++++++++++++++++++++++++++
  
\begin{figure}%++++++++++++++++++++++++++++++++++++++++++++++++++++++
\begin{centering}
	\includegraphics[width=0.4\columnwidth]{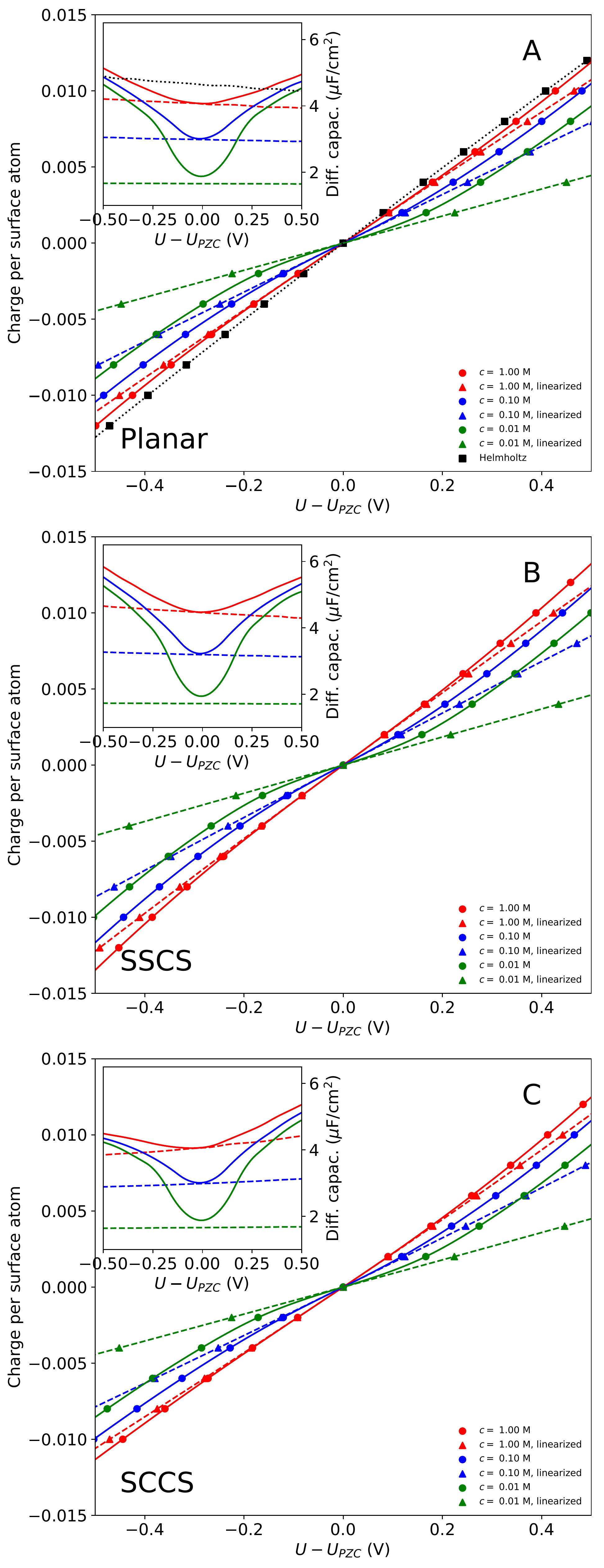}\\
\end{centering}
\caption{Same as Figure \ref{fig:stern-eps01}, but for the full-numerical PB model (circle and solid lines)
and linearized PB model (triangles and dashed lines).
The interface functions employed 
are the following: the planar interface (top, $d=d_0/2 + 6.568$ Bohr, where $d_0$ is the slab thickness, and $\Delta=0.470$ Bohr); the soft-sphere interface
(middle, $r=6.568$ Bohr, $\Delta=0.470$ Bohr); and the SCCS interface (bottom, $\rho_{max}=10^{-4}$ a.u. 
and $\rho_{min}=10^{-5}$ a.u.). The top subplot also includes results obtained with the Helmholtz model
(black squares and dotted line) as a comparison. The same $d$ and $\Delta$ parameters used for the planar
interface have been employed to set the Gaussian countercharge density.}
\label{fig:interface-functions-eps01}
\end{figure}

 Figure \ref{fig:interface-functions-eps01} compares charge-potential curves 
 and the corresponding DC values as computed 
 with the numerical PB solver (Section \ref{subsub:full-pb}) paired to the three different cavities
  presented in Section \ref{sub:cavity}. 
Specifically, we have tested the use of the planar
 interface function also used in Figure \ref{fig:stern-eps01} and \ref{fig:sys-spread-eps01} 
 (see Figure \ref{fig:interface-functions-eps01}A), 
and two additional cavities derived from the SSCS\cite{Fisicaro2017} (Figure \ref{fig:interface-functions-eps01}B)
 and from the SCCS\cite{Andreussi-JCP-2012} (Figure \ref{fig:interface-functions-eps01}C) models, respectively. 
To better compare results across the cavities employed, we choose the corresponding parameters so that
the onsets of the three interface functions lie at approximately the same distance from the metal surface
under neutral conditions, and a similar broadening characterizes the three interfaces.
%For this purpose, we fit the erf-related parameters 
 %in the planar interface function ($d$ and $\Delta$) to the interface function
 %calculated with the SCCS, where we have arbitrarily chosen the parameters $\rho_{max}=10^{-4}$ a.u. 
%and $\rho_{min}=10^{-5}$ a.u.. For the soft-sphere interface, we use the same radius and spread
%parameters as the $d$ and $\Delta$ parameter that have been fitted through the planar interface. 
%We consistently compute similar DC values at the PZC for the three different cavities.

%+++++++++++++++++++++++++++++++++++++++++++++++++++++++

%The computed DC curves for the three interface functions
%under same conditions of bulk electrolyte and applied potential
%are rather similar, consistently with the approximately 
%same distance of the electrolyte-charge onset from the surface. 
Very similar DC-potential curves are obtained using the planar and SSCS
cavities. The former produces slightly higher capacitance values, consistently 
with the electrolyte charge density more closely approaching 
interstitial surface regions with the soft-sphere interface. 
Both interfaces predict DC-potential curves 
that are asymmetric around the PZC, with slightly larger DC values at negative potentials
as compared to the corresponding positive values. 
This is again consistent with the effective separation between the 
surface and the ionic density onset becoming smaller at negative potentials 
due to the larger electron density spilling towards the rigid electrolyte interface. 
Interestingly enough, the trend observed with the SCCS cavity is reversed. 
This density-dependent interface function, in fact, shifts
the ionic density onset further away from the surface as 
the slab charge becomes more negative, effectively increasing the 
electrolyte-slab separation. 

Figure \ref{fig:interface-functions-eps01} also includes results from the 
linearized PB model for the three cavities considered.
This model correctly predicts the DC values at the PZC and the
qualitative DC dependence on the bulk electrolyte concentration.
Note that the capacitance computed with this model
approaches the infinite-screening limit represented by the 
Helmholtz model capacitance for increasing ionic concentrations
(Figure \ref{fig:interface-functions-eps01}A).
As also evident from comparing Figure \ref{fig:stern-linearized-eps01} to Figure \ref{fig:stern-eps01},
the linearized version of the PB model dramatically fails in reproducing the 
potential trend computed with the corresponding non-linear model 
and it returns weak potential dependences with no minimum at the PZC. 
The monotonic trends observed for the linear-regime model 
are consistent with the patterns described for the corresponding non-linear model.
For instance, the density-dependent SCCS cavity predicts monotonically increasing capacitance curves,
as the surface-electrolyte gap increases with increasing potential. This is 
also consistent with the findings of Letchworth-Weaver and Arias\cite{LetchworthWeaver-PRB-2012}, 
who have similarly observed monotonically increasing capacitance curves using a linearized-PB model for the 
diffuse layer and a density-dependent interface function.
%In particular, the DC curve monotonically decreases
%as a function of the potential if a rigid cavity is employed, while it monotonically
%increases if the density-dependent SCCS cavity is used instead. The overall increasing or decreasing 
%trends are consistent with the patterns also observed for the full non-linear PB model. 
 
\begin{figure}%++++++++++++++++++++++++++++++++++++++++++++++++++++++
\begin{centering}
	\includegraphics[width=0.5\columnwidth]{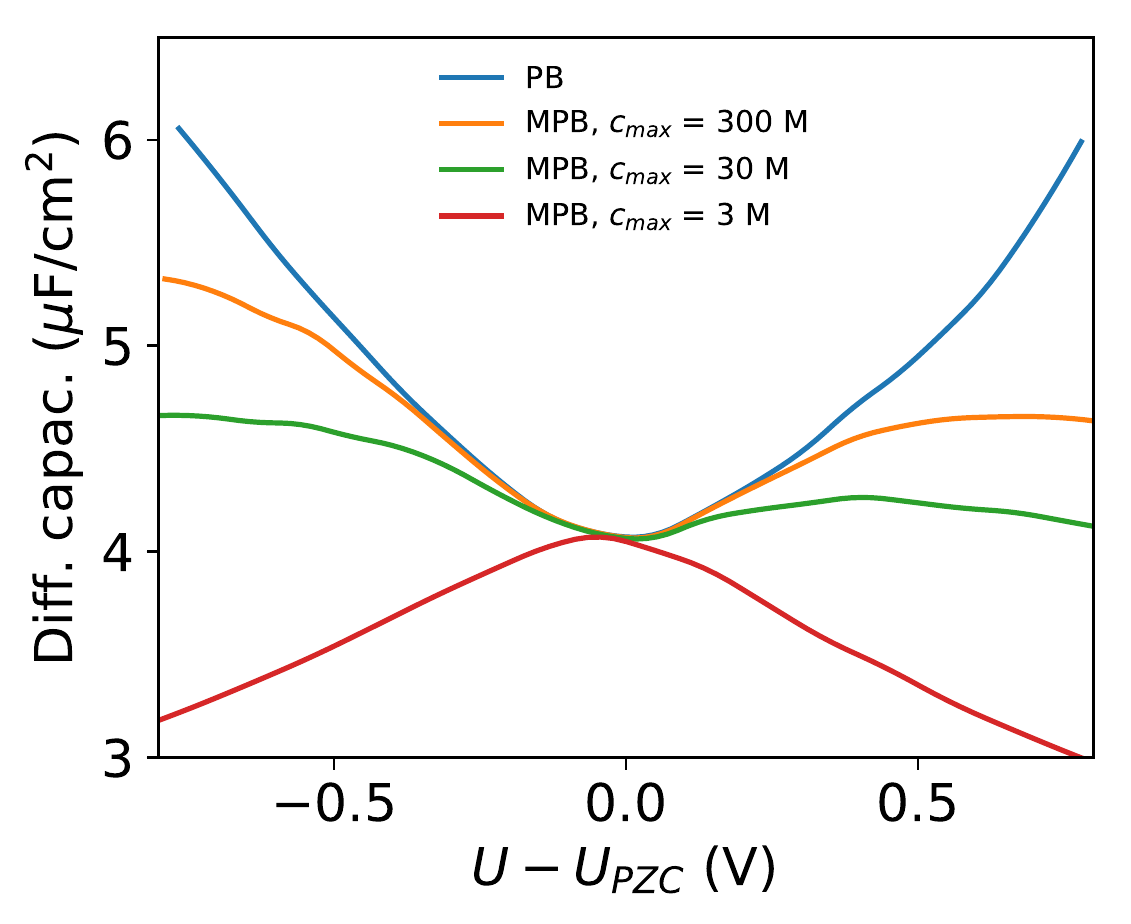}\\
\end{centering}
\caption{The differential capacitance is plotted as a function of the potential for various
values of the $c_{max}$ parameter in the MPB model. The planar interface function 
has been used for all calculations, with $d=6.568$ Bohr and $\Delta = 0.470$ Bohr. }
\label{fig:sys-cmaxs-eps01}
\end{figure}

Figure \ref{fig:sys-cmaxs-eps01} compares DC curves computed using the standard (non-linear) PB model 
to results from the size-modified PB model (see Sections \ref{subsub:MPB} and \ref{subsub:full-pb}). 
In particular, we test values for the $c_{max}$
parameter that range from 300 M to 3 M, which correspond to effective ionic radii from 0.95 \AA\ to 4.39 \AA . 
Introducing a finite size for the ions affects the DC at large
applied potentials: while the standard PB model predicts monotonically increasing
capacitance values for increasing values of the applied potential, the size-modified model 
predicts the DC to first reach a maximum and then decrease as the potential deviates from the PZC. 
The DC maximum is reached at lower values of the potential for decreasing values of the $c_{max}$ parameter.
For $c_{max} = 3$ M the DC-potential curve even 
changes concavity, and the PZC becomes the maximum. 
The observed DC decrease can be explained by the following arguments.
In contrast with the standard PB-model, which allows for infinitely large 
electrolyte concentrations at the interface boundary,
its size-modified variant imposes a maximum local ionic concentration, $c_{max}$. 
When this maximum concentration is locally reached, the steric repulsion between the ions
pushes the ionic charge density
towards the bulk solvent region, effectively increasing the separation between 
the surface and the electrolyte charge, giving rise to the observed DC decrease.

%+++++++++++++++++++++++++++++++++++++++++++++++++++++++

\subsection{Implicit Solvent\label{sub:results-solvent}}

\begin{figure}%++++++++++++++++++++++++++++++++++++++++++++++++++++++
\begin{centering}
	\includegraphics[width=0.5\columnwidth]{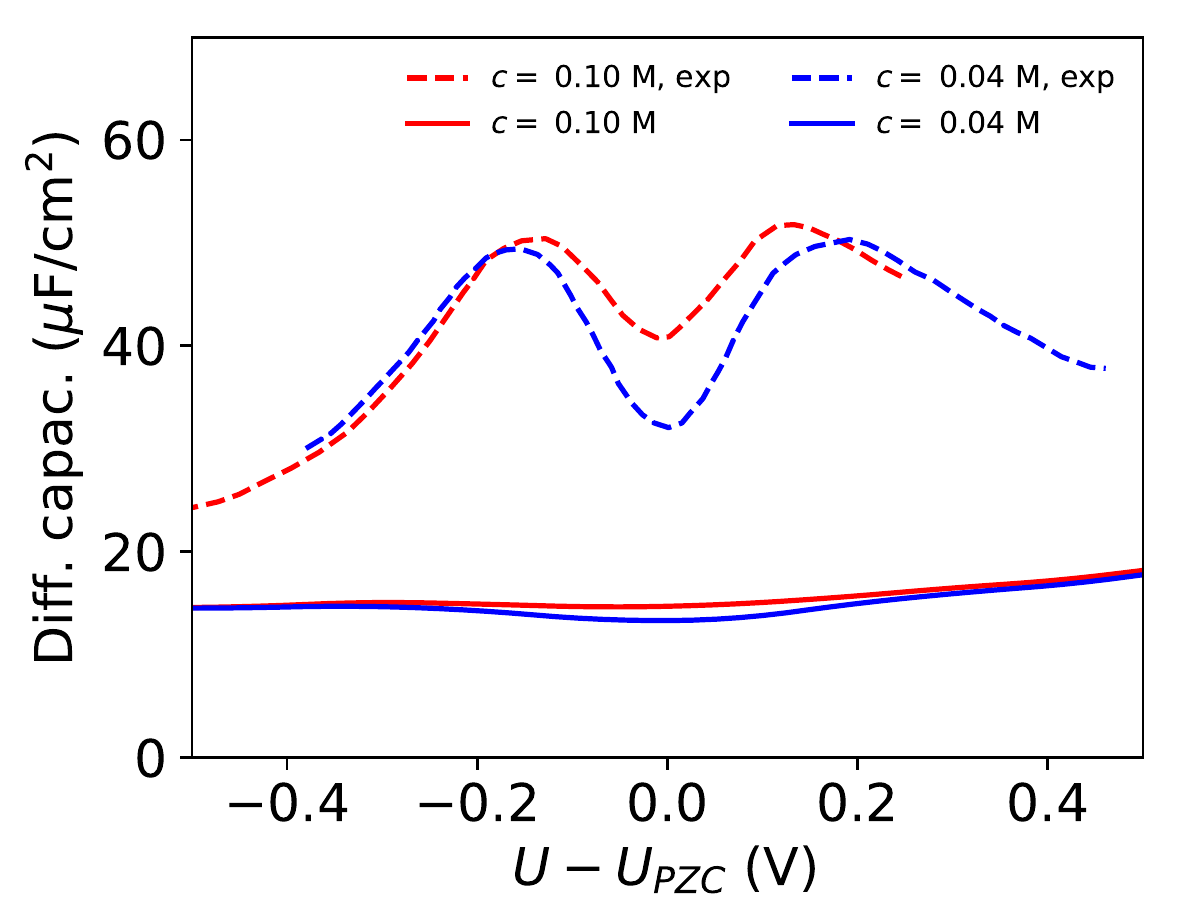}\\
\end{centering}
\caption{The differential capacitance is plotted as a function of the potential. 
Experimental data from Ref.\cite{Valette-JElectroanalChem-1982} are plotted as dashed lines. 
Results of MPB simulations using the SCCS cavity with the original 
parameterization from Ref. \cite{Andreussi-JCP-2012} are plotted as solid lines. 
The value of $c_{max}$ is set to 20 M. Red is for $c_0=0.1$ M and blue is for $c_0 = 0.04$ M. }
\label{fig:sccs}
\end{figure}

After having investigated the performance of the diffuse layer 
models in vacuum, we switch to simulations in implicit water
and compare results to prototypical experimental data, presented in
Figure \ref{fig:sccs}.
In particular, we have considered data reported by Valette\cite{Valette-JElectroanalChem-1982} on the differential 
capacitance of Ag(100) in a KPF$_6$ electrolyte solution.
Consistently with commonly-observed experimental trends, the DC exhibits a `camel-back' shape,
with the minimum indicating the PZC. %Two `humps'  with comparable heights are also observed at approximately $\pm$ 0.16 V (vs PZC). 
As also indicated by Valette, the common potential value at which the potential drop is observed
across the various electrolyte concentrations and the rather symmetric shape of the DC curve around the PZC
suggest a negligible anion adsorption in this electrolyte solution. 
 
 %+++++++++++++++++++++++++++++++++++++++++++++++++++++++
 
On the basis of the results presented in Section \ref{sub:results-vacuum}, 
only the MPB model is expected to qualitatively reproduce
the experimental potential trend, with the capacitance drop at the 
PZC and the DC saturation and decrease at large applied potentials. Instead,
 the Helmholtz and the linearized PB model 
fail to predict the capacitance minimum at the PZC, and the standard PB model 
predicts monotonically increasing capacitance curves. 
Concerning the cavity, the various interface functions 
have been found to give rise to overall similar 
capacitance values in vacuum, at least for parameterizations that lead to similar 
electrolyte charge distributions. The planar and SSCS 
cavities produce essentially identical results, and for this reason in the following
we will consider only the latter, which better suits general interface geometries. 

The DC computed with the MPB model and the SCCS interface including the dielectric continuum
are plotted in Figure \ref{fig:sccs}. We have used here the original SCCS cavity parameters\cite{Andreussi-JCP-2012}, 
which have been fitted to a database of solvation energies of neutral molecules. 
We remind, however, that the non-electrostatic solvation terms have been neglected here. 
Calculations are performed for the experimental bulk electrolyte 
concentrations (0.1 M and 0.04M) and for the steric repulsion 
between ions through the $c_{max}$ parameter, which we 
have initially set to 20 M. Assuming a random close-packing for the ions,
this value of $c_{max}$ corresponds to an effective ionic radius 
of approximately 2.33 \AA . In comparison, experimental upper-bounds for the bare 
(non-solvated) ionic radii are 2.65 \AA\ \cite{Mancinelli-JPCB-2007} 
and 2.42 \AA \cite{Roobottom-JChemEduc-1999}, 
for K$^{+}$ and PF$_{6}^{-}$, respectively. 

As already observed in the vacuum environment, the MPB model 
predicts a DC drop at the PZC, which is more pronounced for the lowest electrolyte concentration
simulated ($c_0$ = 0.04 M). This is in qualitative agreement with measurements. 
However, the overall absolute magnitude of the DC is 
severely underestimated, and the potential dependence computed is also much weaker than in experiments.

\begin{figure}%++++++++++++++++++++++++++++++++++++++++++++++++++++++
\begin{centering}
	\includegraphics[width=0.5\columnwidth]{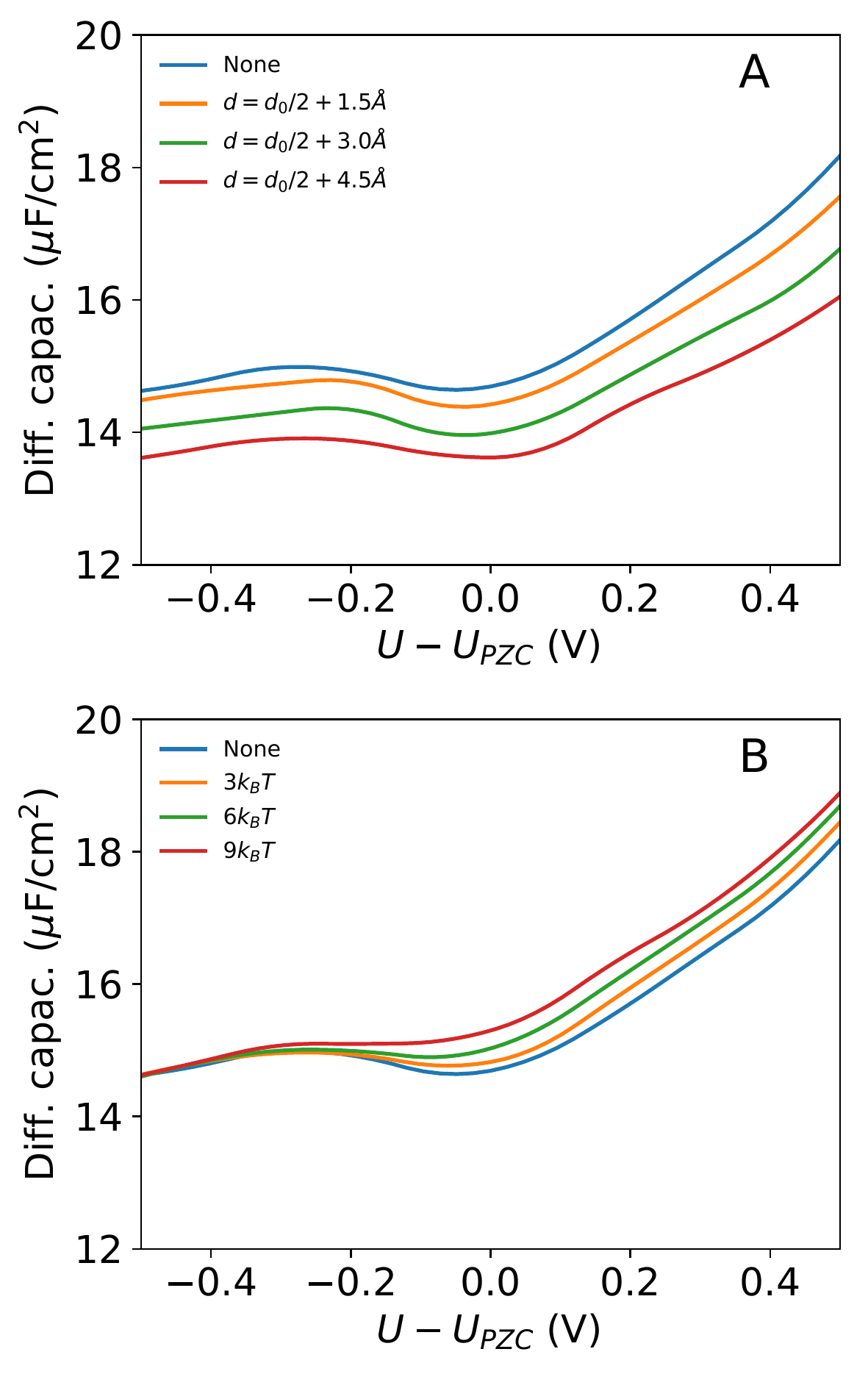}\\
\end{centering}
\caption{The differential capacitance plotted as a function of the potential. 
All data refer to MPB simulations with the SCCS interface. The original
parameterization from Ref. \cite{Andreussi-JCP-2012} has been employed, and $c_0 = 0.1$ M and $c_{max} = 20$ M. 
In the top panel, a repulsive potential between solute and electrolyte is introduced.
Different colors correspond to different values of the $d$ parameter, as indicated ($w$ is set to 0.25\AA ).
In the bottom panel, a Morse-potential interaction between solute and anions is employed instead, with $d = d_0/2 + 1.5$ \AA\ and $w = 0.5$ \AA . 
Different colors correspond to different values of the $E_{ads}$ parameter, as indicated.
  }
\label{fig:sccs-interactions}
\end{figure}

Figure \ref{fig:sccs-interactions} illustrates how the inclusion of additional solute-electrolyte interactions
in the MPB model (see Sections \ref{subsub:additional-interactions}) affects the computed DC (note the different scale of the $y$ axis). 
Figure \ref{fig:sccs-interactions}A shows the effect of including a solute-electrolyte repulsion potential.
This potential introduces a gap between the onset of the dielectric continuum and the 
one of the electrolyte countercharge density. 
In particular, we test various values for the distance parameter $d$ 
in the chosen functional form for the repulsive potential (see Eq. \ref{eq:repulsive-potential}). 
%We set $d = d_0 + \delta$, where $d_0$ is half the slab thickness and $\delta$ an 
%additional spacing, which we vary in the range 1.5 \AA\ to 4.5 \AA . 
By setting the spread parameter $w$ to 0.25 \AA\ we ensure a fast decay of the exponential repulsion. 
The effect of introducing this Stern-layer gap is essentially a rigid shift of the DC curve.
The observed capacitance decrease is consistent with the corresponding 
increase of the surface-electrolyte charge distance for increasing $d$ values, 
and the small magnitude of the shift is related to the large dielectric constant 
that characterizes the region where the electrolyte charge is located.

%+++++++++++++++++++++++++++++++++++++++++++++++++++++++

Figure \ref{fig:sccs-interactions}B illustrates how the DC is affected 
by anion adsorption as accounted through the continuum model of Baskin and Prendergast\cite{Baskin-JElectrochemSoc-2017}. 
For the solute-anion interactions, we set the following values for the Morse-potential parameters: 
$w = 0.5$ \AA\ and $d = d_0/2 + 1.5$ \AA , where $d_0$ is the slab thickness. The adsorption energy $E_{ads}$ is varied 
in a range from $3k_BT$ to $9k_BT$ (i.e. from 80 meV to 230 meV at 300 K). 
Note that we simulate the asymmetric anion adsorption without accounting for solute-cation interactions. 
At the most negative potentials considered, where the electrostatic attraction
of cations is much stronger than the imposed solute-anion interaction, the anion adsorption
does not alter the computed DC. At the highest potentials simulated, the additional 
attractive interaction between surface and anions increases the electrolyte 
countercharge at the interface, thereby increasing the DC. 
At intermediate potentials, the anion adsorption 
shifts the DC minimum from the PZC towards negative potentials, 
where the electrostatic interaction compensates the 
anion attractive potential. 

\begin{figure}%++++++++++++++++++++++++++++++++++++++++++++++++++++++
\begin{centering}
	\includegraphics[width=0.5\columnwidth]{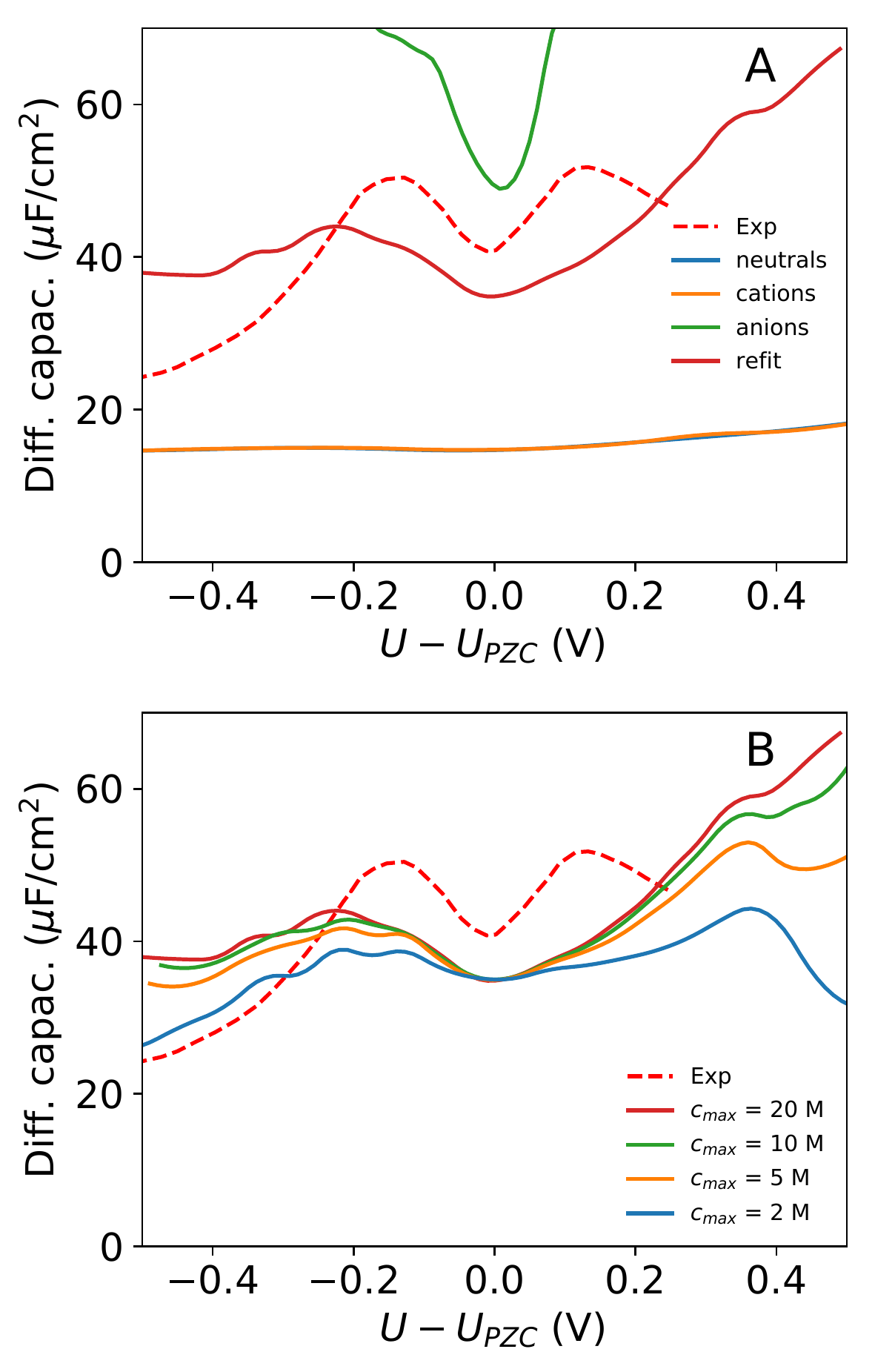}\\
\end{centering}
\caption{The differential capacitance is plotted as a function of the potential for $c_0 = 0.1$ M. 
Experimental data\cite{Valette-JElectroanalChem-1982} are shown as dashed lines. 
All theoretical data refer to MPB simulations with the SCCS interface function. 
In the top panel, results from the original SCCS parameterization \cite{Andreussi-JCP-2012} (`neutral'),
are compared to results from the cation- and anion-specific parameterizations\cite{Dupont-JCP-2013} and to results
from the parameter fit to the Pt(111) PZC\cite{Hoermann-JCP-2018}. The value of $c_{max}$ is set to 20 M.
 In the bottom panel, the SCCS parameterization from Ref.\cite{Hoermann-JCP-2018} is employed, and the 
value of $c_{max}$ varied from 20 M to 2 M, as indicated.}
\label{fig:sccs-parametrization}
\end{figure}

The cavity parameterization is found to have a much
 larger influence on the absolute value of the computed DC. 
This is illustrated in Figure \ref{fig:sccs-parametrization}, where we plot the DC-potential curves calculated 
with the original interface parameterization that was optimized for neutral isolated systems
and with the two parameterizations that have been later proposed\cite{Dupont-JCP-2013} to best fit 
anion and cation solvation energies, respectively. The DC computed with the cation-specific 
parameterization does not significantly differ from the one obtained from the 
original parameterization. This is consistent with the very similar values for the cavity parameters
$\rho_{max}$ and  $\rho_{min}$ in the two fits. Significantly different cavity parameters
were instead found to best fit the anion database, and we consistently observe a considerable
difference in the resulting capacitance. 
In particular, the anion-specific parameterization is characterized by smaller cavities, and the reduced 
gap between the surface and the continuum fluids give rise to larger DC values. 
As illustrated in Figure \ref{fig:sccs-interactions}A, the spacing between
the surface and the electrolyte countercharge density has a rather contained effect on the 
absolute DC value. These findings suggest that the gap between the electrode surface 
and the dielectric polarization charge is thus the main responsible for the large DC 
dependence on the cavity parameterization, as also suggested by Sundararaman et al. \cite{Sundararaman2018ImprovingCalculations}.  
Consistently, Melander \emph{et al.}\cite{Melander2018}, who have employed a 
dielectric cavity based on the van der Waals radii of the surface atoms, have found 
that increasing the atomic radii leads to significantly lower capacitance values.
%This is consistent with the effective dielectric constant in the former spacing 
%being about 80 times smaller than the corresponding value in the latter spacing. 

%+++++++++++++++++++++++++++++++++++++++++++++++++++++++

It is evident that none of the three 
SCCS cavities described so far is able to describe well experimental data: 
the original SCCS parameterization and the cation refit underestimate the measured DC, while the anion parameterization  
overestimates it. This is not surprising, considering that all three parameter sets have been 
fitted to solvation energies of isolated systems. Figure \ref{fig:sccs-parametrization}A also 
includes a DC curve computed with cavity parameters that have been recently 
fitted\cite{Hoermann-JCP-2018} to reproduce  the theoretical estimate of the absolute PZC of Pt(111). 
This last paramerization overall returns the best agreement 
with experimental data, even though it underestimates the measured DC in the potential region 
close to the PZC and more severely overestimates it for larger applied potentials. 

As clearly shown from the simulations in vacuum, the capacitance of the MPB model at large 
absolute potentials is strongly affected by the steric-repulsion between the ions (cf. Figure \ref{fig:sys-cmaxs-eps01}).
Figure \ref{fig:sccs-parametrization}B shows the effect of decreasing the value of $c_{max}$ 
from 20 M to 2 M, which is equivalent to increasing the ionic particle radii from from 2.33 \AA\ to 5.02 \AA . 
Decreasing $c_{max}$ broadens the minimum in correspondence of the PZC and 
lower the capacitance at the highest and lowest potentials examined,
improving agreement with experimental data. 
It is thus tempting to suggest that 
effective radii larger that the bare ones should 
be employed for the electrolyte particles, as also suggested in the literature on the 
basis of the strongly-bound solvent molecules 
surrounding ions in solution\cite{Bazant2009}. 
Note that experimental estimates for the radius of the solvated K$^{+}$ ions 
span a range \cite{Bazant2009} from 3.8 \AA\ \cite{Gering2006} to 6.62 \AA \cite{Nightingale1959}, which would largely justify the range of $c_{max}$  
investigated.

\begin{figure}%++++++++++++++++++++++++++++++++++++++++++++++++++++++
\begin{centering}
	\includegraphics[width=0.5\columnwidth]{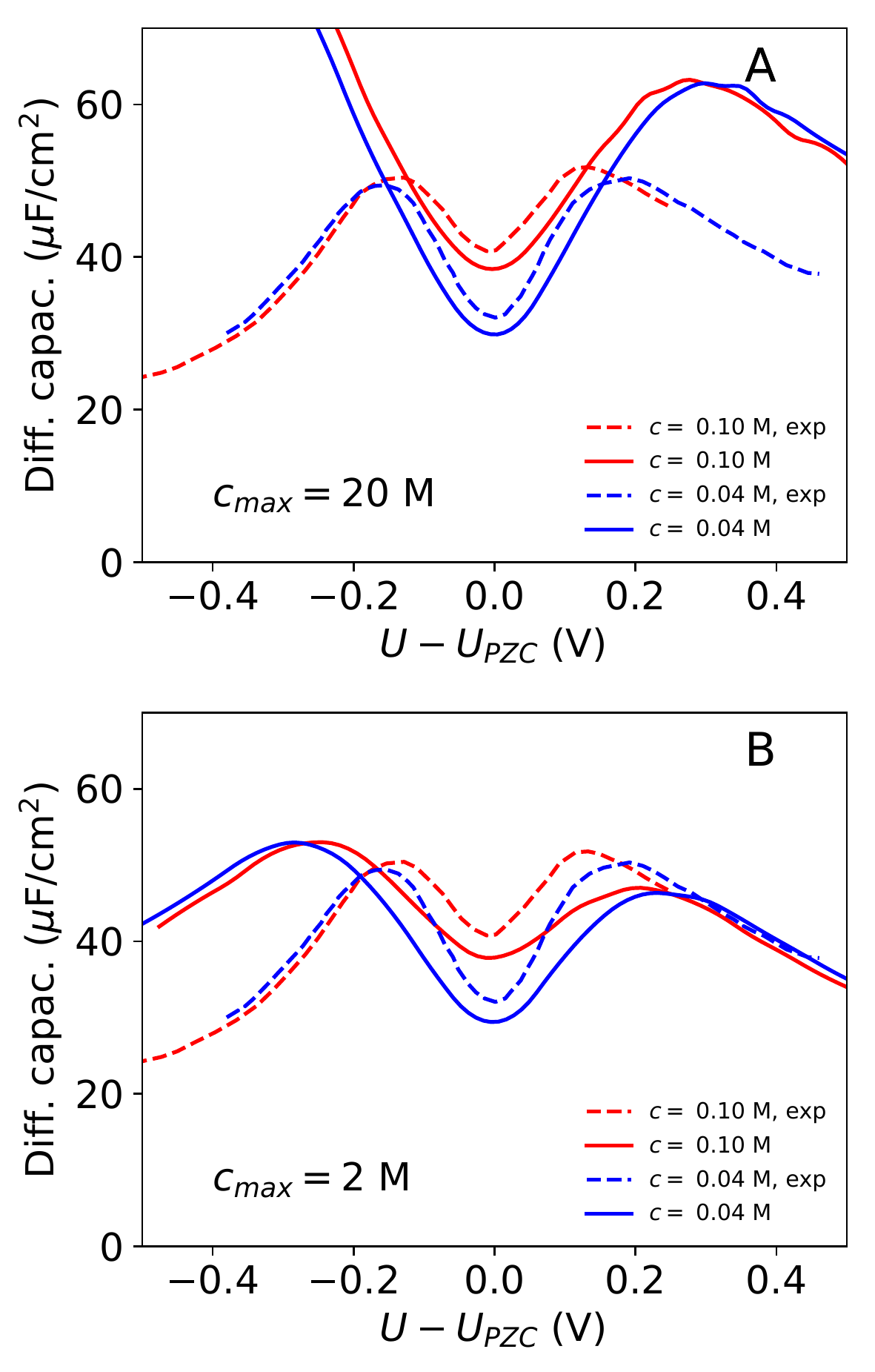}\\
\end{centering}
\caption{Same as Figure \ref{fig:sccs}, but the SSCS interface
has been employed in the simulations. In top and bottom panel the value of $c_{max}$ is set to 20 M and 2 M, respectively.}
\label{fig:soft-spheres}
\end{figure}

After having investigated the DC capacitance computed using density-based cavities, 
we now consider simulations performed with the SSCS interface function. 
Results are presented in Figure \ref{fig:soft-spheres}, where we have used the parameterization proposed 
by Fisicaro et al.\cite{Fisicaro2017}. 
%In particular, the cavity size is determined by  the Ag ionic radius as tabulated in
% the unified force-field (UFF) times a scaling factor $f = 1.16$. 
Figure \ref{fig:soft-spheres}A reports the capacitance computed using the upper-bound $c_{max}$ value of 20 M. 
Despite the cavity %multiplication factor $f$ 
parameters
were originally fitted to solvation energies
of isolated systems as for the SCCS interface, the SSCS model leads to a very good 
description of the experimental DC around the PZC for the two electrolyte 
concentrations considered. As also observed for the SCCS interface function,  
the capacitance at large absolute potentials is instead overestimated when using $c_{max} = 20$ M.
Similar to the case of vacuum (Section \ref{sub:results-vacuum}), the (M)PB model
returns asymmetric DC curves with the rigid cavity from the SSCS model. This finding can again be explained on the basis 
of the extent by which the electron density spilling from the metal surface approaches the continuum. 
The separation between the surface and the electrolyte onset, in fact, is effectively reduced 
for lower values of the potential, with a subsequent increase of the capacitance values.

%+++++++++++++++++++++++++++++++++++++++++++++++++++++++

Figure \ref{fig:soft-spheres}B shows the computed DC curve for the lower
$c_{max}$ value of 2 M ($r_i = 5.02$ \AA).  
The agreement with experiments is 
significantly improved, and both the position and the 
height of the `humps' are comparable to measured data.
Thus, also calculations performed with the SSCS cavity 
suggest that ionic radii larger than the bare ones should be employed to 
limit the steric crowding at electrode interfaces. 
Note that 
Sundararaman et al. recently achieved a similarly good description of the DC of Ag(100)
with a soft-sphere-based continuum model\cite{Sundararaman2018ImprovingCalculations}.
The cavity size in their model was also based on the Ag ionic radius as tabulated in
the unified force-field (UFF)\cite{rappe_jacs_1992}, times a scaling constant. As also noted by Sundararaman et al. \cite{Sundararaman2018ImprovingCalculations}, 
the good description of the Ag(100) DC might be thus inferred to the Ag UFF ionic radius 
being fortuitously suitable to describe the cavity size for this system. 
Future investigations of the DC for other systems will shed light on this point.
%This is not surprising: our findings suggest that the computed DC curves heavily depends on 
%the size of the continuum cavity (cf. Figure \ref{fig:sccs-parametrization}). This, in turn, is to large extend determined 
%by the tabulated ionic radii for the involved atomic species (Ag). 
%This could be a fortuitous event, especially considering that the remaining parameters in these 
%soft-sphere approaches were fitted to solvation energies of isolated systems
%\cite{Fisicaro2017, Sundararaman2018ImprovingCalculations}.

Despite the overall good agreement with experimental data, 
the DC predicted using the SSCS cavity overestimates and underestimates the measured
data at negative and positive potentials, respectively. 
The trend observed with the soft-sphere cavity in implicit solvent is consistent with the trends observed in vacuum with rigid cavities, which
are found to predict an overall decreasing DC with increasing potential. 
In comparison, measurements exhibit slightly larger capacitance values 
at positive potentials, in better agreement with trends observed with the
density-dependent SCCS cavity. 
Future work will clarify whether improved agreement with experimental data
can be achieved by a specific refitting of the SCCS cavity.

\begin{figure}%++++++++++++++++++++++++++++++++++++++++++++++++++++++
\begin{centering}
	\includegraphics[width=0.5\columnwidth]{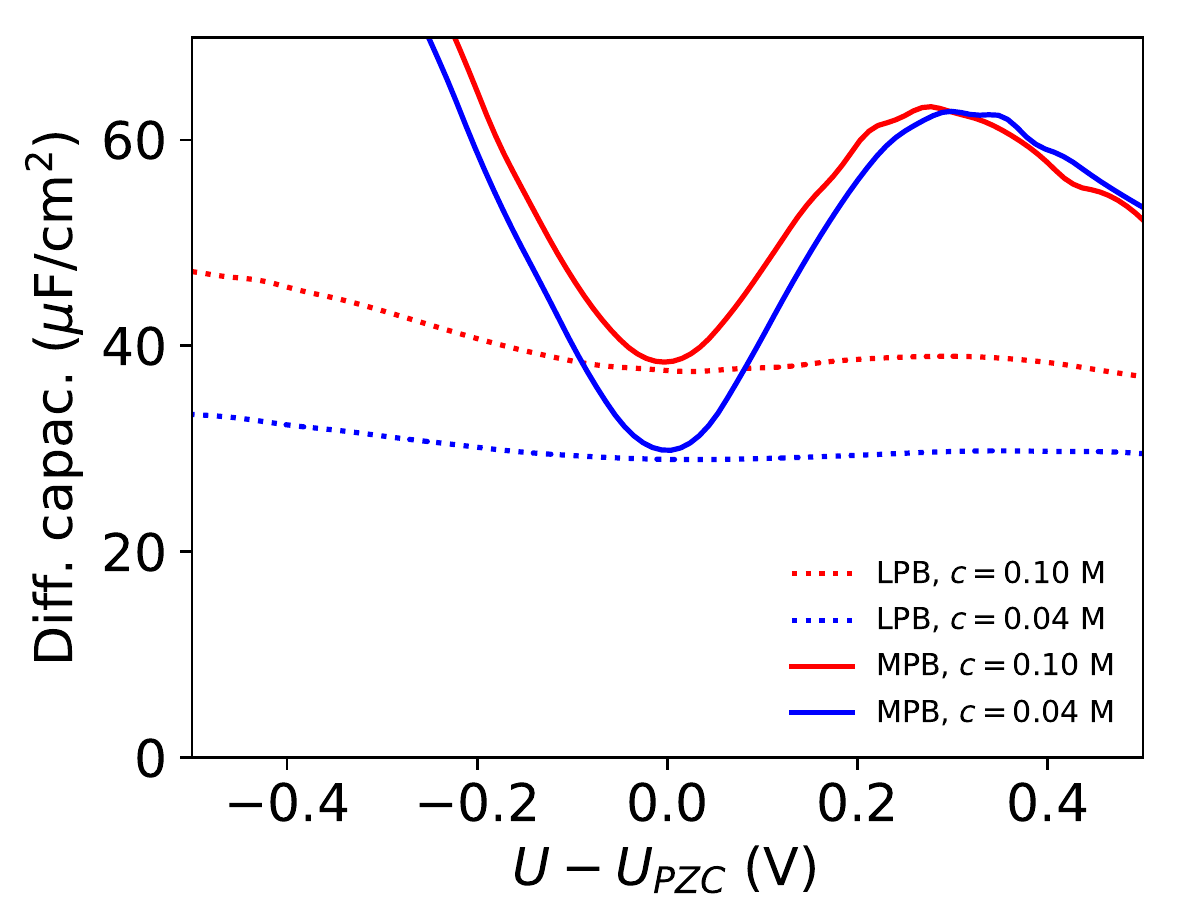}\\
\end{centering}
\caption{Same as Figure \ref{fig:soft-spheres}, but the DC curves obtained 
with the MPB model and  $c_{max} = 20$ M are compared to analogous curves obtained
with the linearized PB model.}
\label{fig:LPB-MPB}
\end{figure}

Regardless on the cavity employed, our findings suggest that the MPB model for the electrolyte
is able to capture the main features of the experimental DC for Ag(100) in an ideally non-adsorbing ionic solution. 
We note in passing that in addition of being more physically sound, the MPB model 
is also more numerically stable than the standard PB model, as the extremely large ionic charge densities 
that the latter predicts at the boundary between the electrified surface 
and the solvent region are difficult to handle with the numerical 
solvers without the inclusion of a Stern layer. 
While avoiding such instabilities, the linearized-PB model is inadequate for 
describing the capacitance of a charged metal surface. As expected from the results in vacuum (Section \ref{sub:results-vacuum}),
Figure \ref{fig:LPB-MPB} illustrates how the linear-regime model predicts essentially potential-independent capacitance values, 
which are only accurate close to the PZC.

Our results are in contrast with the finding from Ref. \cite{Sundararaman2018ImprovingCalculations}, where an additional
non-linear dielectric model was suggested to be necessary in order to reproduce 
the trends observed in the measurements. 
Our findings also differ from the ones of Melander \emph{et al.}\cite{Melander2018},
who have reported potential-independent capacitance trends for a metal surface (Au(210)) 
in an electrolyte solution using both a linearized- and a non-linear- MPB model. 
While discrepancies with Refs. \cite{Sundararaman2018ImprovingCalculations, Melander2018} require further investigation, 
our results are consistent with data from full-continuum models \cite{Bazant2009, Nakayama2015, Baskin-JElectrochemSoc-2017}, 
where the solution of the non-linear MPBE is found to lead to the experimentally-observed `camel-back'-shape for the 
DC curve for metal surfaces in aqueous solutions.

%+++++++++++++++++++++++++++++++++++++++++++++++++++++++

%%%%%%%%%%%%%%%%%%%%%%%%%%%%%%%%%%%%%%%%%%%%%%%%%%%%%%%%%%%%%%%%%%%%%%
\section{SUMMARY AND CONCLUSIONS\label{sec:Summary-and-Conclusions}}

In summary, we have presented a hierarchy of electrolyte models 
that can be integrated in the framework of DFT to account for the 
presence of the diffuse layer in first-principles simulations of electrochemical interfaces. 
We have validated the accuracy of the models by 
comparing computed DC values to experimental data, 
focusing on the Ag(100) surface in an aqueous electrolyte as study system.

Results suggest that the size-modified PB model is necessary in order 
to reproduce the main characteristics of the experimental DC, i.e.
the concentration-dependent drop at the PZC and the two 
local maxima at intermediate applied potentials. The lowest-rung  
planar Helmholtz model, which does not include any dependence
on the bulk electrolyte concentration, predicts
negligible DC dependences on the applied potential. Similarly, the standard PB
model, both in the linear-regime and in its full non-linear implementations
fails in describing experimental DC trends. 

Further accounting for solvent effects through a continuum 
dielectric allows for a direct comparison
of computed DC values to experimental data.
We observe a large influence of the choice of the dielectric
cavity on the absolute DC values, consistently with previous findings \cite{Sundararaman2017EvaluatingImprovement, Sundararaman2018ImprovingCalculations}.
For the SCCS interface function, the best agreement with experimental 
data is obtained for a parameterization of the cavity that is fitted\cite{Hoermann-JCP-2018}
to reproduce an interface-related observable, i.e. the 
theoretical estimate of the PZC of Pt(111). 
The original parameterization of the SSCS  cavity 
has been found instead to produce a relatively 
good agreement 
with experimental data without the need of refitting.

While it is important to stress how the different approaches can be extended and tuned to improve the description of electrochemical systems, it is worth pointing out that the reported analysis is based on continuum models that only account for part of the physical phenomena occurring at electrified interfaces. At the center of the reported analysis is the description of the diffuse layer, of its shape and characteristics. Nonetheless,  a more realistic model should account for the different sizes of the ions composing the electrolyte. Moreover, as it is also clear from the results reported, the dielectric properties of the liquid solution at the interface with a solid substrate need to be properly modeled in order for the continuum approach to be meaningful. The bare substrate and, even more, a charged interface will induce order and rigidity in the overlaying liquid, substantially affecting the dielectric permittivity over a distance of one or more solvation layers. The fact that current state-of-the-art continuum models are not able to describe with the same accuracy systems with different charge states, and in particular require a separate parameterization for anions, is clearly a limitation of the current techniques in dealing with electrified interfaces. Similarly, non-linear effects in the dielectric response of the liquid may account for some of the deviations observed at higher applied potentials. Other possibly minor effects that are not explicitly accounted for in the presented models are the ones related to the change in dielectric screening of the electrolyte solution for high concentrations of the diffuse layer. Cancellation of errors resulting from the parameterization of the model may lead to an approach that seems accurate, but lacks transferability. As more and more ingredients are added and carefully tuned, they will unlock the full potential of continuum models for electrochemical setups. 

%%%%%%%%%%%%%%%%%%%%%%%%%%%%%%%%%%%%%%%%%%%%%%%%%%%%%%%%%%%%%%%%%%%%%%
% Acknowledgements
\begin{acknowledgments}
This project has received funding from the European Union's Horizon 2020 research and innovation programme under grant agreements No. 665667 and No. 798532. 
N. M. and O. A. acknowledge partial support from the MARVEL National Centre of Competence in Research of the Swiss National Science Foundation.
This work was supported by a grant from the Swiss National Supercomputing Centre (CSCS) under project ID s836.
Part of the computational developments presented in this work have been pursued during the UNT Hackathon 2018, sponsored by the ORAU Envent Sponsorship Program and the University of North Texas.
\end{acknowledgments}

%%%%%%%%%%%%%%%%%%%%%%%%%%%%%%%%%%%%%%%%%%%%%%%%%%%%%%%%%%%%%%%%%%%%%%
% Bibliography
\bibliographystyle{aipnum4-1}{}
%\bibliography{bibliography}
\bibliography{Mendeley}

\end{document}